\documentclass[aps,prl,twocolumn,superscriptaddress,floatfix,nofootinbib,longbibliography]{revtex4-2}
\usepackage{amsthm}
\usepackage{amsmath,amssymb,color,comment,physics}
\usepackage[makeroom]{cancel}
\usepackage[caption=false]{subfig}
\usepackage{mathrsfs}
\usepackage{graphicx}
\usepackage{subfig}
\usepackage[countmax]{subfloat}
\usepackage[english]{babel}
\usepackage{dsfont}
\usepackage[normalem]{ulem}
\usepackage[bookmarks=true,colorlinks,linkcolor=OrangeRed,urlcolor=NavyBlue,citecolor=RoyalBlue]{hyperref}
\usepackage[dvipsnames]{xcolor}
\usepackage{braket}

\usepackage{color}

\newcommand{\blue}[1]{{\color{blue}{#1}}}

\begin{document}

\title{Tunable Topological Phases in Multilayer Graphene Coupled to a Chiral Cavity}

\author{Sayed Ali Akbar Ghorashi$^+$}\email{sayedaliakbar.ghorashi@stonybrook.edu}
\affiliation{Department of Physics and Astronomy, Stony Brook University, Stony Brook, New York 11794, USA}

\author{Jennifer Cano}
\affiliation{Department of Physics and Astronomy, Stony Brook University, Stony Brook, New York 11794, USA}
\affiliation{Center for Computational Quantum Physics, Flatiron Institute, New York, New York 10010, USA}

\author{Ceren~B.~Dag$^+$}
\email{ceren\_dag@g.harvard.edu}

\affiliation{Department of Physics, Indiana University, Bloomington, Indiana 47405, USA}
\affiliation{ Department of Physics, Harvard University, Cambridge, Massachusetts 02138, USA}
\affiliation{ITAMP, Harvard-Smithsonian Center for Astrophysics, Cambridge, MA 02138, USA}

\begin{abstract}
Coupling photonic cavity fields to electronic degrees of freedom in 2D materials introduces an additional control knob to the toolbox of solid-state engineering. Here we demonstrate a subtle competition between cavity frequency and interlayer tunneling in graphene stacks that is responsible for topological phase transitions in light-matter Hilbert space and that cannot be captured by mean-field theory in vacuum. A systematic exploration of multilayer graphene heterostructures and stacking configurations in a chiral tHz cavity reveals that linear dispersion enhances the low-energy cavity-induced topological gap. Furthermore, in bilayer graphene, a displacement field drives the low-energy vacuum band from valley-Chern to Chern insulator, comprising a gate-tunable topological phase transition.
Our findings pave the way for future control and engineering of graphene heterostructures with chiral cavity fields.
\end{abstract}

\maketitle

\def\thefootnote{+}\footnotetext{These authors contributed equally to this work.}

\blue{\emph{Introduction}}.
Cavity-QED engineering of quantum materials has emerged as a powerful tool to manipulate phases of matter \cite{hubener2021engineering,schlawin2022cavity,bloch2022strongly,RevModPhys.91.025005,lu2025cavityengineeringsolidstatematerials}.
In a typical experimental setup, materials are not driven by external fields, but instead are strongly or ultra-strongly coupled to enhanced vacuum fluctuations in a cavity, modifying material properties such as electronic transport, energy gaps, or phase transitions \cite{li2018vacuum,paravicini2019magneto,thomas2021large,doi:10.1126/science.abl5818,jarc2022cavity,herzig2024high,kipp2024cavityelectrodynamicsvander}.
At the same time, due to their exceptional tunability, 2D materials have garnered significant attention as platforms to realize and control phenomena such as electronic transport, superconductivity, and topology \cite{Novoselov_2016,review2d,Ren_2016}. Thus, it is highly desirable to apply cavity QED engineering to 2D materials, thereby introducing a new tuning knob with the potential to realize new phases of matter \cite{nguyen2023electronphoton,PhysRevB.109.195173,PhysRevLett.132.166901}.

Chiral cavities present a new degree of freedom whereby time-reversal symmetry can be controllably broken.
In general, a 2D material in a circularly polarized cavity couples to both right- and left-circularly polarized photons.
A difference in the frequency or light-matter coupling of the polarizations results in a chiral cavity that breaks the time-reversal symmetry \cite{PhysRevB.110.L121101}. Several schemes for chiral cavities have recently been proposed \cite{hubener2021engineering,tay2024terahertzchiralphotoniccrystalcavities} and realized \cite{suarez2024chiral,PhysRevB.109.L161302,suarezforero2024chiralquantumopticsrecent}.
They have been predicted to induce time-reversal broken topological phases~\cite{wei2024cavityvacuuminducedchiralspinliquids,PhysRevLett.132.166901}, including the Chern insulator phase in a graphene monolayer \cite{PhysRevB.84.195413,PhysRevB.99.235156,PhysRevB.107.195104,PhysRevB.110.L121101}.

\begin{figure}[t!]
    \centering
    \includegraphics[width=0.9\linewidth]{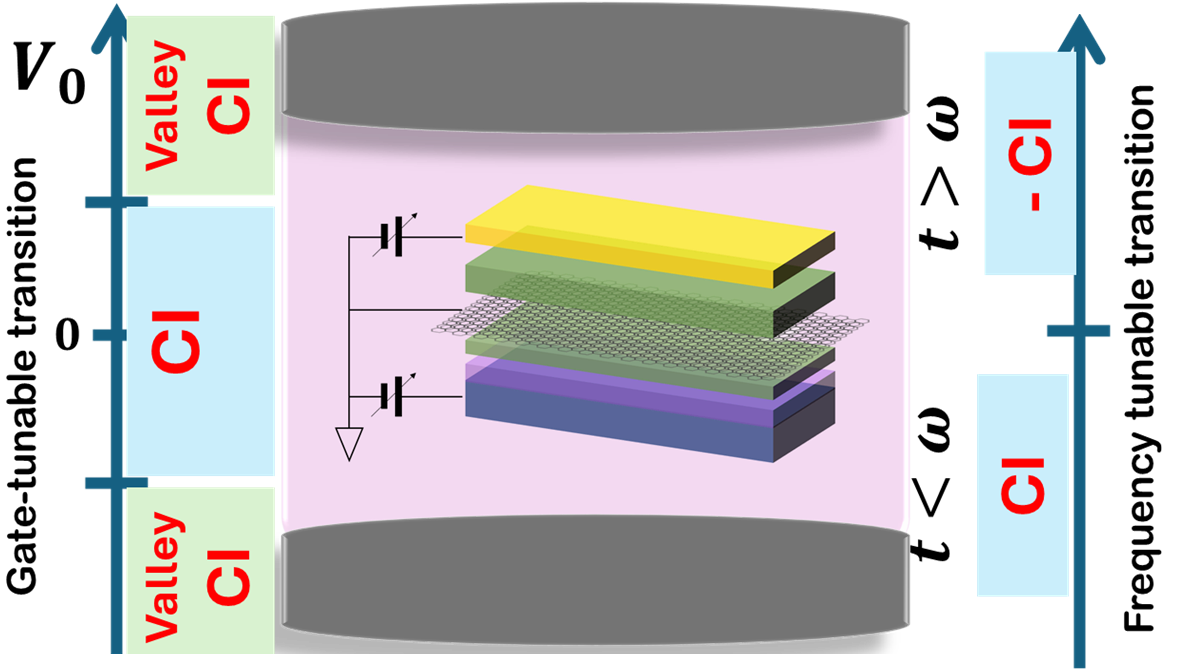}
    \caption{\textbf{Multilayer graphene coupled to a chiral cavity field.} We observe two topological transitions: (i) between two Chern insulators (CI) with opposite Chern numbers $|C|$ by tuning the interlayer coupling (or cavity frequency); and (ii) between a CI and valley CI by tuning displacement field.}
    \label{fig:adpic}
\end{figure}

Here we present the first systematic study of cavity-QED engineering in multilayer graphene heterostructures coupled to a chiral cavity.
Our results are summarized in Fig.~\ref{fig:adpic}.
In contrast to monolayer graphene, the extra degree of freedom afforded by the interlayer tunneling gives rise to a rich phase diagram.
Specifically, we uncover topological phase transitions (TPTs) in bilayer, trilayer, and tetralayer graphene coupled to a chiral cavity, which result from a competition between the interlayer tunneling and the cavity frequency. While mean field theory (MFT) in vacuum reveals time-reversal and chiral symmetry breaking, as in monolayer graphene, it does not access these TPTs.

Introducing a tunable displacement field, which explicitly breaks the inversion symmetry of the heterostructure, allows for a distinct and gate-tunable TPT between valley-Chern and Chern insulators. The gate-tunable phase transition is captured by the MFT only in the regime where the interlayer tunneling is lower than the cavity frequency. Furthermore, we find hybrid light-matter band topology, as was also found in a graphene monolayer \cite{PhysRevB.110.L121101} and transition metal dichalcogenides \cite{nguyen2023electronphoton} coupled to the chiral cavity. Interestingly, however, even-layer stacks of graphene reveal a band above the vacuum, with zero Chern number, yet accompanied by chiral edge modes. We show that this arises from the Berry curvature at topological light-matter hybridization points.

Importantly, we determine that
all multilayer structures with dispersion beyond linear exhibit a cavity-induced gap much smaller than that of the monolayer graphene, except for ABA Bernal trilayer graphene, which also has a linear branch. Therefore, a linear dispersion is observed to be favorable for a larger cavity-induced gap in graphene stacks.

\begin{figure*}[htb!]
\centering
\includegraphics[width=1.9\columnwidth]{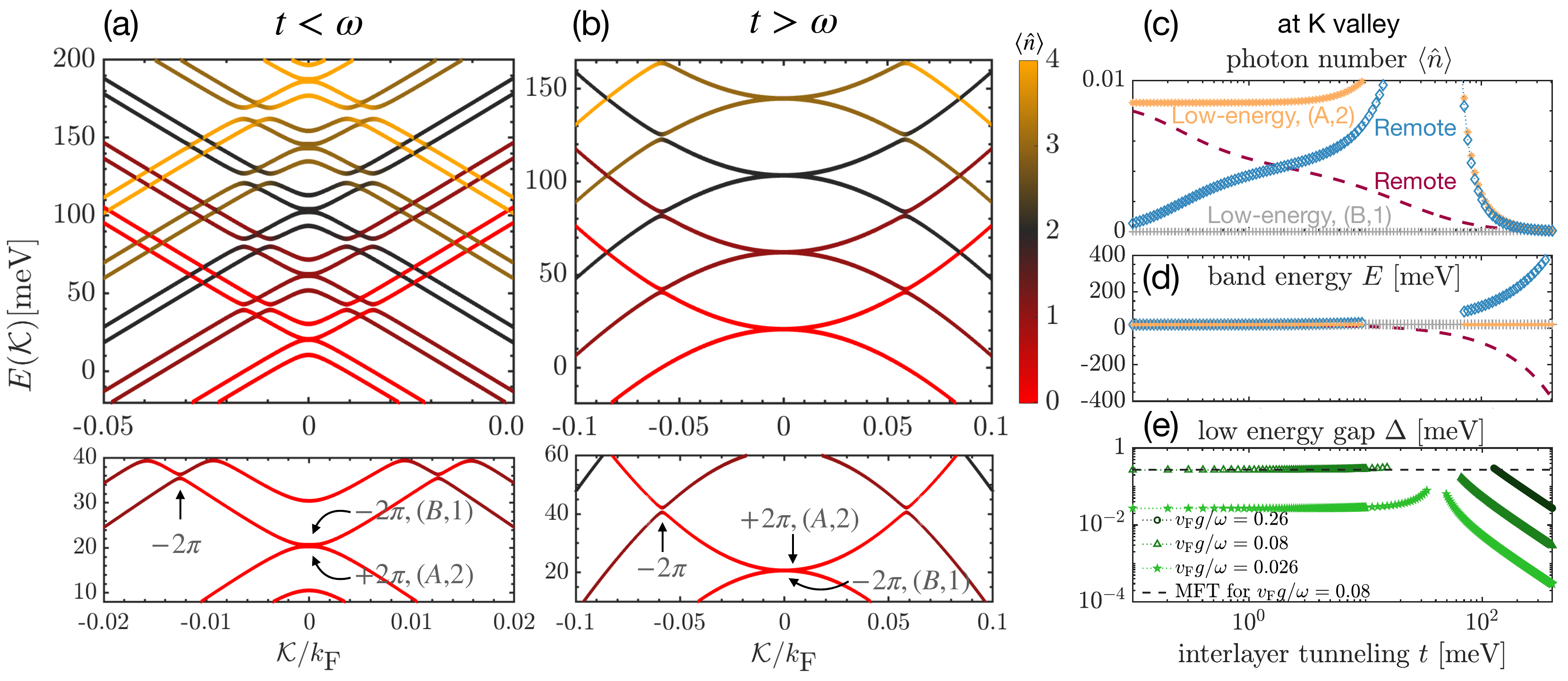}
\caption{\textbf{Band structure and the vacuum gap} (a)-(b) Hybrid band structure of the $\mathbf{K}$ valley of bilayer graphene in a chiral cavity with frequency $\omega_c=10$ THz and strong light-matter interaction strength $v_{\rm F} g/\omega_c = 0.09$ at zero displacement field when the interlayer coupling is (a) less than the cavity frequency, e.g.,~$t=10$meV and (b) larger than the cavity frequency, e.g.,~$t=400$meV. Remote bands are not visible in (b). Lower panels focus on the vacuum bands; the color of the band denotes its photon number. Berry phases at electronic and photon-exchange hybrid avoided crossings are marked in the lower panels. The band structure is exactly the same at the opposite valley contributing the same Berry phase. (c) Photon number (d) band energy and (e) low-energy gap of vacuum bands at $\mathbf{K}(\mathbf{K'})$ point with respect to interlayer tunneling for $\omega_c=10$ THz and $V_0=0$. The light-matter interaction strength $v_{\rm F} g/\omega_c = 0.09$ is fixed in (c)-(d), whereas it changes in (e). Gap decreases with decreasing light-matter interaction or increasing interlayer tunneling.}\label{Fig1}
\end{figure*}

\blue{\textit{Light-matter interactions.}}~We consider a chiral cavity with a single circularly polarized mode, which can be engineered from a magnetoplasma in, e.g., InSb sandwiched between Fabry-Perot cavity mirrors \cite{tay2024terahertzchiralphotoniccrystalcavities}. The quantized vector potential is given by, ${\mathbf{\hat A}}=\sqrt{\frac{1}{\epsilon_0\mathcal{V}2\omega_c}}\left[\mathbf{e}_R \hat {a}^{\dagger}+\mathbf{e}_R^* \hat {a}\right],$
where $\mathbf{e}_R=(1,\textrm{i})/\sqrt{2}$ is the polarization vector and $\mathcal{V}=\chi \left(2\pi c/ \omega_{c}\right)^3 $ \cite{paravicini2019magneto} is the effective cavity volume with the light concentration parameter $\chi$. Here, the operators $[ a,a^{\dagger} ] = 1$ are the circularly polarized photon operators renormalized by the diamagnetic term, and hence the cavity frequency is renormalized as $\omega=\sqrt{\omega_c^2+\omega_D^2}$ where $\omega_c$ is the frequency of the bare cavity and the diamagnetic frequency $\omega_D=e/\sqrt{m\epsilon_0 \mathcal{V}}$ with $m=1$ the electron mass~\cite{PhysRevB.110.L121101}. The coupling of vacuum fluctuations to bilayer graphene can be described by the continuum Hamiltonian (with $\hbar=e=1$),
\begin{eqnarray}
\hat H(\mathbf{k})  &=&  \hat \psi^{\dagger}(\mathbf{k}) \bigg\lbrace v_{\rm F} \tau^0 \left[ \xi (k_x- \hat A_x) \sigma^1 +(k_y- \hat A_y )\sigma^2\right]     \label{eq:fullHamiltonian}\\
&+&\frac{t}{2}(\tau^1\sigma^1-\tau^2\sigma^2) +V_0 \tau^3\sigma^0 \bigg\rbrace\hat \psi(\mathbf{k})  + \omega\left( {\hat a}^{\dagger} {\hat a}+\frac{1}{2}\right),\notag
\end{eqnarray}
where $\hat \psi^{\dagger}(\mathbf{k})=\left(\hat c^{\dagger}_{A\mathbf{k}1},\hat c^{\dagger}_{B\mathbf{k}1}, \hat c^{\dagger}_{A\mathbf{k}2}, \hat c^{\dagger}_{B\mathbf{k}2}\right)$, and $\hat c_{r\mathbf{k}l}$ annihilate electronic degrees of freedom on the graphene sublattice $r=A,B$ and layer $l$. $v_{\rm F}$ is the Fermi velocity of graphene, $t$ is the interlayer tunneling, $V_0$ is the displacement field and $\xi=\pm1$ denotes the valley. The Pauli matrices $\tau$  and $\sigma$ act on layer and sublattice, respectively. The light-matter interaction is described by
\begin{eqnarray}
    \hat H_{\mathbf{K},\rm int}(\mathbf{k}) &=& - v_{\rm F} g \sum_{l} \hat a^{\dagger} \hat c^{\dagger}_{A\mathbf{k}l}\hat c_{B\mathbf{k}l}  + \rm h.c.  \label{eq:interactionH}
\end{eqnarray}
 Eq.~\eqref{eq:interactionH} simply adds a layer degree of freedom to the analogous term for monolayer graphene \cite{PhysRevB.110.L121101}. The interaction Hamiltonian at the other valley is obtained by the mapping $\hat H_{\mathbf{K'},\rm int}(\mathbf{k}) = - \hat H_{\mathbf{K},\rm int}^{\dagger}(\mathbf{k},A \leftrightarrow B)$.
The light-matter coupling amplitude can be written in terms of microscopic parameters as $g = \frac{\alpha}{m}\sqrt{2\pi/(\mathcal{V}\hspace{.5mm}\omega})$ \cite{PhysRevB.110.L121101,tay2024terahertzchiralphotoniccrystalcavities}, where $\alpha=2.68$ a.u.~is the lattice spacing.

\begin{figure}[htb!]
\centering
\includegraphics[width=1\columnwidth]{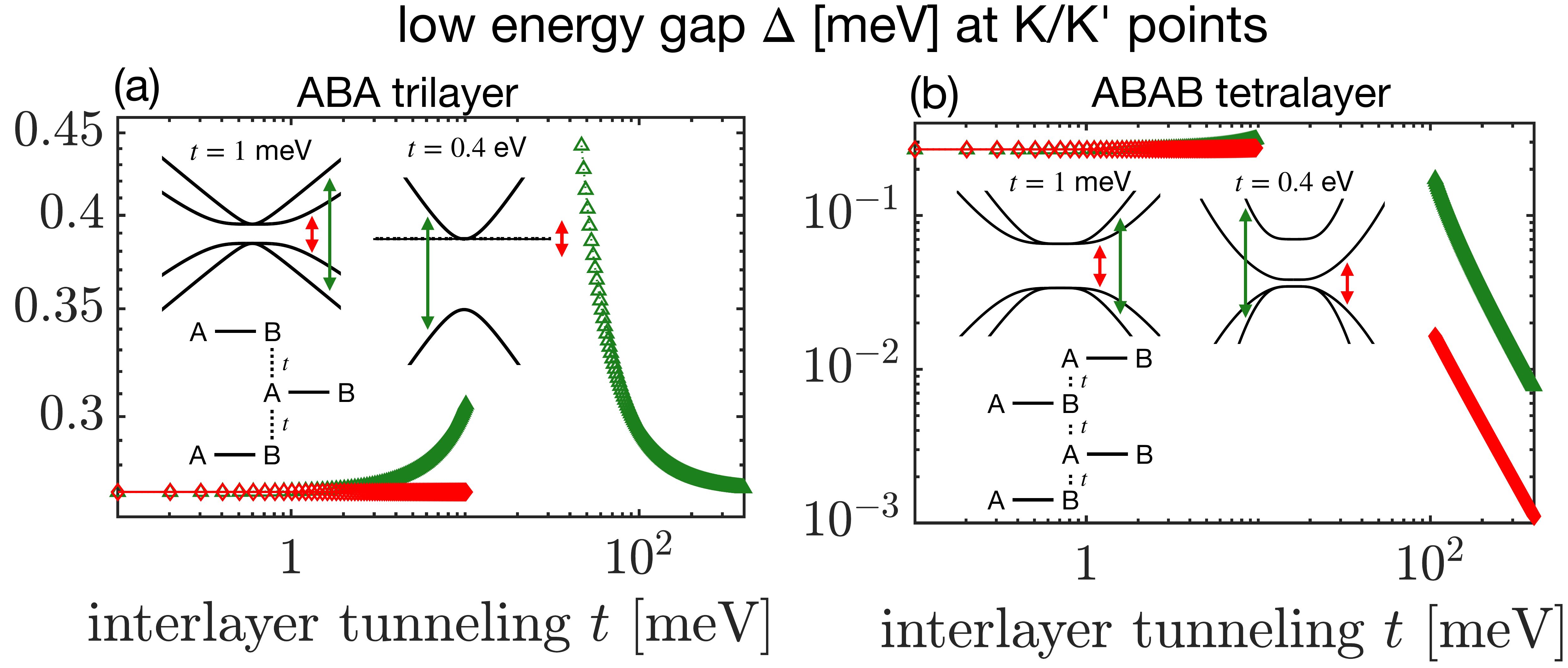}
\caption{\textbf{Low energy gaps at $\mathbf{K(K')}$ valleys for Bernal-stacked multilayer graphene with respect to the interlayer tunneling $t$ at $v_{\rm F} g/\omega\sim0.08$}. In both (a) ABA trilayer and (b) ABAB tetralayer graphene, we define two different gaps denoted by red and green arrows in the inset band structures. Insets also show the atomic configurations. Bernal-stacked trilayer shows an enhanced vacuum gap for large $t$ (green-triangles in (a)), a striking difference from bilayer and chiral-stacked multilayers.}\label{Fig4}
\end{figure}

To gain insight into the influence of the cavity on bilayer graphene we derive the leading vacuum projected ($\langle \hat a^{\dagger}\hat a\rangle \equiv \langle \hat n\rangle =0$) cavity mediated interactions via a Schrieffer-Wolff procedure \cite{PhysRevB.110.L121101}:
\begin{eqnarray}\label{interaction}
 &\mathcal{\hat H}_{\mathbf{K},\rm vac} =- \frac{ v_{\rm F}^2 g^2}{\omega}\sum_{\mathbf{k} \mathbf{k'}}\Bigg( \sum_{m} \hat c^{\dagger}_{B\mathbf{k'}m}\hat c_{A\mathbf{k'}m}\hat c^{\dagger}_{A\mathbf{k}m}\hat c_{B\mathbf{k}m} \notag \\
    &+ c^{\dagger}_{A\mathbf{k'},2}c_{B\mathbf{k'},2}c^{\dagger}_{B\mathbf{k},1}c_{A\mathbf{k},1} + c^{\dagger}_{A\mathbf{k'},1}c_{B\mathbf{k'},1}c^{\dagger}_{B\mathbf{k},2}c_{A\mathbf{k},2}\Bigg).
\end{eqnarray}
Note that $\mathcal{\hat H}_{\mathbf{K'},\rm vac}=\mathcal{\hat H}_{\mathbf{K},\rm vac}(A\leftrightarrow B)$. The first term in Eq.~(\ref{interaction}) is an intralayer interaction, identical in form to the same term for monolayer graphene \cite{PhysRevB.110.L121101}. The other terms describe interlayer interactions new to multilayer systems.
In principle, these term can lead to interlayer order parameters such as $\braket{\hat c^{\dagger}_{A\mathbf{k'}1}\hat c_{A\mathbf{k}2}}$.
However, to preserve rotational symmetry, which our numerics indicate is preserved, these terms must be \(k\)-dependent and vanish at \(k=0\). Here, because we are interested in the physics around the $\mathbf{K}(\mathbf{K'})$ point, we leave the effect of these channels for future studies. Instead, we focus on the mean-field description of Eq.~\eqref{interaction}, $\mathcal{\hat H}_{\rm vac}^{\rm mft}(\mathbf{k})=\hat \psi^{\dagger}(\mathbf{k})\hat h_{\rm vac}^{\rm mft}(\mathbf{k}) \hat \psi(\mathbf{k})$, with the following intralayer order parameters which depend on the valley index $\xi$,
\begin{align}\label{mft}
    \hat h_{\rm vac}^{\rm mft}(\mathbf{k})=&- \frac{v_{\rm F}^2 g^2}{ \omega}\begin{bmatrix}
       \Delta_1  &  0 & 0 & 0\\
       0  &   \Delta_2 & 0 & 0 \\
       0 & 0 & \Delta_3 & 0 \\
       0 & 0 & 0 & \Delta_4
    \end{bmatrix},
\end{align}
where, $\Delta_1=-\langle \hat c^{\dagger}_{B\mathbf{k},1}\hat c_{B\mathbf{k},1} \rangle - (\xi-1)/2,\, \Delta_2=-\langle \hat c^{\dagger}_{A\mathbf{k},1}\hat c_{A\mathbf{k},1} \rangle + (\xi+1)/2,\, \Delta_3=-\langle \hat c^{\dagger}_{B\mathbf{k},2}\hat c_{B\mathbf{k},2} \rangle  - (\xi-1)/2,\, \Delta_4=-\langle \hat c^{\dagger}_{A\mathbf{k},2}\hat c_{A\mathbf{k},2} \rangle + (\xi+1)/2$.
At $V_0 = 0$ and any $t$, our self-consistent calculation  yields $\hat h_{\rm vac}^{\rm mft}(\mathbf{k})=- \frac{v_{\rm F}^2 g^2}{2 \omega} \left(\xi \tau^0 \sigma^3 -\tau^0 \sigma^0  \right)$, written for both valleys \cite{supp}. This reveals time-reversal symmetry breaking in bilayer graphene along with breaking chiral symmetry.
The inversion symmetry of the bilayer at $V_0=0$ is preserved \cite{supp}.
These observations are expected to hold more generally for all chiral-stacked multilayer graphene.

We now compare the mean-field Hamiltonian in Eq.~\eqref{mft} to numerical diagonalization of Eq.~\eqref{eq:fullHamiltonian} to study the vacuum bands in cavity-coupled bilayer graphene.

\begin{figure}[htb!]
\centering
\includegraphics[width=0.9\columnwidth]{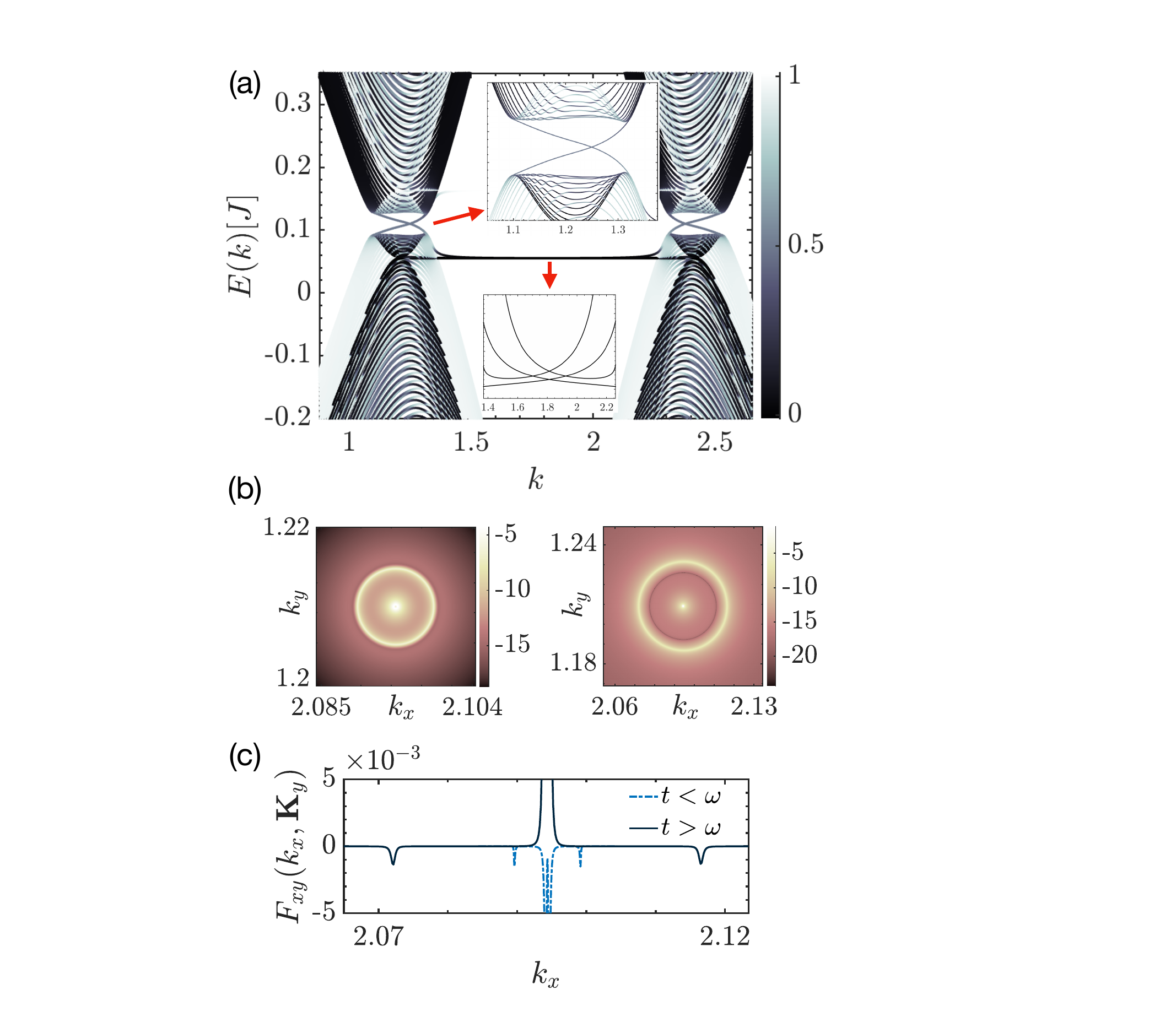}
\caption{\textbf{Chiral edge modes at zero displacement field and hybrid band topology} (a) Tight-binding band structure for a bilayer graphene slab with zigzag edges, coupled to a chiral cavity mode in the regime $t>\omega$ with $t=0.4J$ and $\omega=0.11J$ where $J=1$  is fixed for computational convenience. Colorbar denotes the photon number of the bands. Lower inset: 4 chiral electronic edge modes connecting two valleys exhibiting time-reversal symmetry breaking. Upper inset: 4 chiral hybrid edge modes with nonzero photon number, 2 at each valley traversing the topological light-matter gap. (b) Berry curvature of the first low-energy hybrid band, $|F_{xy}(k_x,k_y)|$ for $t < \omega$, e.g.,~$t=10$meV (left) and $t > \omega$, e.g.,~$t=400$meV (right), both at $\omega_c=10$THz. (c) A cross-section for these two cases showing the sign of Berry curvature. For $t < \omega$ the hybrid band has Chern number $-4$, whereas at $t > \omega$ the Chern number vanishes.}\label{Fig2}
\end{figure}
\blue{\textit{Interlayer tunneling induced TPT}}.~ We first turn off the displacement field ($V_0=0$) and study the competition between interlayer tunneling $t$ and cavity frequency $\omega$ in the regimes $t < \omega$ and $t > \omega$. When $t<\omega$, the two graphene layers are only weakly coupled. Nevertheless, as we explain, the physics differs from the monolayer. At $t=0$ the spectrum consists of two doubly degenerate linear dispersions. Finite $t$ lifts the degeneracy, yielding two low-energy quadratic bands that touch only at the $\mathbf{K}(\mathbf{K'})$ point. The hybrid light-matter band structure obtained by numerically diagonalizing Eq.~\eqref{eq:fullHamiltonian} is shown in Fig.~\ref{Fig1}(a), where the colors denote the photonic population of each band. The lowest four bands in the vicinity of $\mathbf{K}(\mathbf{K}')$ points are vacuum-like, i.e.,~$\braket{\hat n} \approx 0$, which reveal a small cavity-induced gap between bands with opposite orbital/layer character. The low-energy vacuum gap is approximately captured by MFT in Eq.~\eqref{mft} for a sufficiently low light-matter interaction, e.g.,~$v_{\rm F} g/\omega \leq 0.1$, as shown in Fig.~\ref{Fig1}(e). This is because the MFT predicts no dependence of this gap on the interlayer tunneling, whereas the gap is numerically calculated to have a slight dependence on it. The cavity-induced topological gap generates a $2\pi$ winding at both Dirac points leading to $C=2$ including both valleys for the low-energy vacuum band. The latter is defined as the band with no light-matter hybridizations, e.g.,~the band with $(A,2)$ orbital character at $\mathbf{K}$ point in the lower panel of Fig.~\ref{Fig1}(a).

\begin{figure}[htb!]
\centering
\includegraphics[width=1\columnwidth]{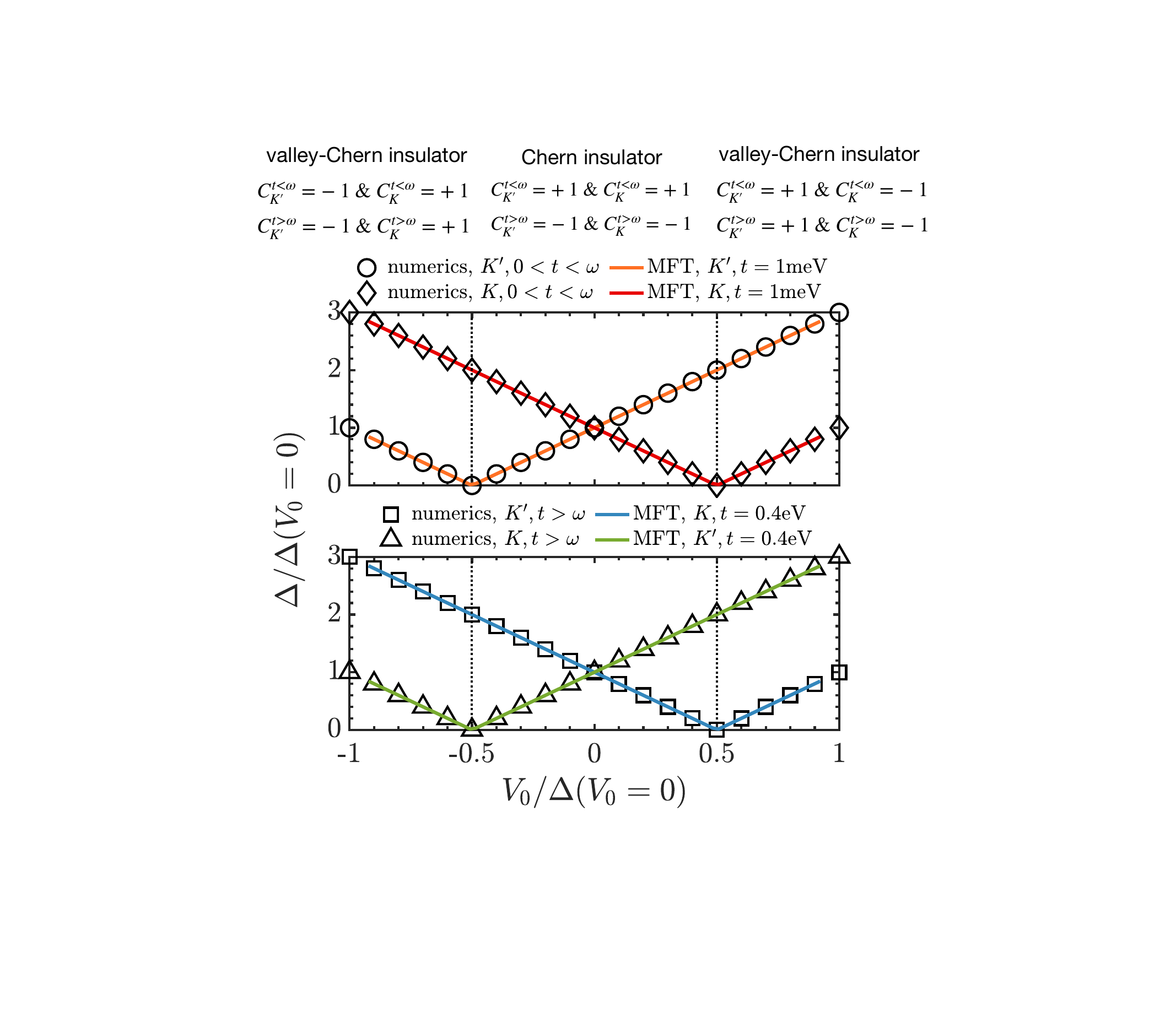}
\caption{\textbf{Topological phase transitions driven by displacement field $V_0$ at $0<t < \omega$ and $t > \omega$}.  In both regimes, the gap closes at either valley, also at displacement field $V_0=\pm \Delta(V_0=0)/2$ regardless of the choice of parameters, driving the system between valley-Chern and Chern insulator phases. The MFT is accurate only for $t < \omega$, whereas for $t > \omega$ it predicts the opposite behavior for the two valleys.}\label{Fig3}
\end{figure}
In the second regime, $t>\omega$, which is the realistic regime for multilayer graphene in a tHz cavity, the low-energy bands are well separated from the remote bands (Fig.~\ref{Fig1}(b) and Fig.~\ref{Fig1}(d)). As the interlayer tunneling increases, the vacuum gap decreases as a power-law in $t$, Fig.~\ref{Fig1}(e). This gap also decreases with decreasing light-matter interaction.

In the regime in which the two energy scales $t$ and $\omega$ are comparable, we observe significant photon fluctuations in these low-energy bands, leading to an absence of a vacuum band (see SM \cite{supp}). Although there is always one low-energy band that retains its vacuum character in the vicinity of the $\mathbf{K}$ point, i.e.,~$\braket{\hat n}=0$, the other acquires a finite number of photons, $\braket{\hat n} \gg 0$ in this transition regime, Fig.~\ref{Fig1}(c). This is why we cannot define a vacuum gap at $t \sim \omega$ (the blank region in Fig.~\ref{Fig1}(e)). The photon fluctuations in these low-energy bands are caused by a process in which the remote bands with higher photon population move downward, hybridizing with them. This process, importantly, changes the orbital character of the low-energy vacuum band, once it exists, from $(A,2)$ to $(B,1)$, as shown in the lower panels of Fig.~\ref{Fig1}(a)-(b), also altering the Chern number of this band from $C=+2$ at $t<\omega$ to $C=-2$ at $t>\omega$.

Since the low-energy vacuum band in regimes $t<\omega$ and $t>\omega$ is predominantly electronic, its Chern number can be probed in electronic transport measurements \cite{PhysRevB.110.L121101}.
Although we have described the transition by increasing $t$, it is more likely to be observed by changing the frequency of the cavity in fixed interlayer tunneling, which can be achieved with movable mirrors~\cite{jarc2022cavity,digiorgio2023surface}.
Since the analytical method integrates out the photonic degrees of freedom and is projected to vacuum, it is oblivious to these transitions occurring in the light-matter Hilbert space, and hence is inaccurate at $t>\omega$.

It is straightforward to generalize these results to multilayer graphene. We compare trilayer and tetralayer graphene with (i) chiral-stacking (ABC and ABCA) and (ii) Bernal-stacking (ABA and ABAB) \cite{min2008electronic}.
Our numerical results indicate that the cavity-induced gaps in the chiral-stacked trilayer (cubic dispersion) and tetralayer (quartic dispersion) are similar to those shown for the bilayer. However, the story changes for Bernal-stacked trilayer graphene,
which has linear and quadratically dispersing bands that behave differently under cavity coupling.
Fig.~\ref{Fig4}(a) shows how linear (green) and quadratic (red) gaps change with interlayer tunneling (see inset in Fig.~\ref{Fig4}(a)). The linear gap is lower bounded by $v_{\rm F}^2 g^2/\omega \approx 0.27$ meV (the graphene monolayer gap), while the quadratic gap is zero for $t > \omega$. On the other hand, Bernal-stacked tetralayer behaves quite similarly to that of bilayer and chiral-stacked multilayers, Fig.~\ref{Fig4}(b). The dispersion consists of two quadratic branches and the gaps between each quadratic band decay as a power-law in the regime $t > \omega$, similar to bilayer. Thus, we conclude that the linear dispersion favors a larger gap in multilayer graphene coupled to a chiral cavity at $t>\omega$.

All multilayers undergo a TPT in the light-matter Hilbert space at $t \sim \omega$, as detailed above for bilayer graphene.

\blue{\textit{Low-energy hybrid bands and chiral edge modes.}}
Going beyond the vacuum band in bilayer graphene, in the regime $t <\omega$, the first low-energy hybrid band winds opposite to the vacuum band at their avoided crossing at $\mathbf{K}$ point
(lower panel of Fig.~\ref{Fig1}(a)). Note that we call a band `hybrid' when it exhibits a light-matter avoided-crossing, e.g.,~the band above the vacuum band in Fig.~\ref{Fig1}(a). Unlike the vacuum band, this band receives another contribution from the avoided crossing around the 1-photon exchange gap (see the Berry curvature in the left panel of Fig.~\ref{Fig2}(b)) where the photonic and electronic degrees of freedom are entangled \cite{PhysRevB.110.L121101}, causing another winding of $-2\pi$. Thus, the first low-energy hybrid band features $C=-4$, a hybrid Chern number \cite{nguyen2023electronphoton,PhysRevB.110.L121101}.
The change in the topology of the vacuum band with interlayer interaction as discussed in the previous section also has an important consequence for the topology of the low-energy hybrid band. As the winding due to the topological light-matter hybridizations remains unchanged, reversing the sign of the contribution from the Dirac point (Fig.~\ref{Fig1}(b) lower panel) compensates for the former and as a result leads to a $C=0$ band in the regime $t > \omega$. Despite the vanishing Chern number, the Berry curvature of the low-energy hybrid band exhibits a local structure due to the Dirac node and topological light-matter hybridizations, the right panel of Fig.~\ref{Fig2}(b) and Fig.~\ref{Fig2}(c). Indeed Fig.~\ref{Fig2}(a) shows a tight-binding simulation revealing chiral edge states \cite{McCann_2013,supp}. Specifically, the gap at the light-matter hybridization point here hosts chiral edge modes, which have both electronic and photonic character (upper inset).

Furthermore, we find four electronic edge modes originating from vacuum bands with two crossings (lower inset).
The chiral edge modes split when $t\neq 0$, which is beyond the vacuum-projected MFT in Eq.~\eqref{mft}.

\blue{\textit{Gate-tunable TPT.}}~In the absence of a cavity, a displacement field breaks inversion symmetry in bilayer graphene and induces a gap of the same sign per valley. However, as discussed above, the cavity-induced gap alters its sign in each valley. Therefore, tuning $V_0$ must drive a gap-closing phase transition at either $\mathbf{K}$ or $\mathbf{K'}$, depending on the sign of $V_0$, to a valley-Chern insulator, Fig.~\ref{Fig3}. The MFT matches accurately with the numerics at $t < \omega$, but fails at $t > \omega$ as expected. In the specific case of uncoupled double-layer graphene, i.e.,~\( t = 0 \), and displacement field $V_0=0$, the bands are doubly degenerate. As \( V_0 \) is varied, the degeneracy is lifted, and both valleys undergo a phase transition simultaneously at the same critical values of \( V_0=\pm \frac{\Delta(V_0=0)}{2} \) \cite{supp}. However, unlike \( t \neq 0 \), this only results in a sign change in the Chern number that leads to the absence of a valley-Chern insulator. We expect all chiral stacked multilayers to exhibit this gate-tunable TPT. The same is not true for Bernal-stacked multilayers where inversion symmetry is already broken in the absence of a displacement field.

\blue{\emph{Discussion and Outlook.}}~We focused primarily on direct interlayer hopping $t$ and physics near the Dirac points. However, remote hopping in multilayer graphene can significantly impact electronic correlations and low-energy gaps, e.g., by inducing van Hove singularities due to trigonal warping of the Fermi surface away from the Dirac nodes \cite{McCann_2013,min2008electronic}. Because trigonal warping is a momentum-dependent term, the cavity couples to the warping term, resulting in a light-matter interaction of $H_{w}^{\rm int} \propto -\sqrt{2} g \hat a^{\dagger}\hat c^{\dagger}_{A\mathbf{k}1}\hat c_{B\mathbf{k}2} + \textrm{h.c.}$. Therefore, trigonal warping would enhance the cavity-induced electronic interactions between two layers, which we leave for future investigations.

The subtle competition between interlayer hopping and cavity frequency highlighted in this work calls for an investigation into the interplay between interlayer Coulomb and cavity-induced interactions on an equal footing. Furthermore, in systems with an enhanced density of states, exploring the competition between electron-electron and cavity-induced interactions presents an exciting research direction \cite{PhysRevB.109.195173,PhysRevLett.131.176602,PhysRevLett.132.166901,PhysRevLett.130.196201,PhysRevB.107.195423}.

\textit{Acknowledgments.}--We thank V. Karle, L. Chen for discussions. C.B.D acknowledges support from the NSF through a grant for ITAMP at Harvard University. This research was supported in part by grant NSF PHY-2309135 to the Kavli Institute for Theoretical Physics (KITP).
S.A.A.G. and J.C. acknowledge support from the Air Force Office of Scientific Research under Grant No. FA9550-20-1-0260. J.C. acknowledges additional support from the Alfred P. Sloan Foundation through a Sloan Research Fellowship, and from the Flatiron Institute, a division of the Simons Foundation. The authors gratefully acknowledge support from the Simons Center for Geometry and Physics, Stony Brook University at which some of the research for this paper was performed.

\bibliographystyle{apsrev4-2}
\bibliography{main.bib}

\begin{thebibliography}{34}%
\makeatletter
\providecommand \@ifxundefined [1]{%
 \@ifx{#1\undefined}
}%
\providecommand \@ifnum [1]{%
 \ifnum #1\expandafter \@firstoftwo
 \else \expandafter \@secondoftwo
 \fi
}%
\providecommand \@ifx [1]{%
 \ifx #1\expandafter \@firstoftwo
 \else \expandafter \@secondoftwo
 \fi
}%
\providecommand \natexlab [1]{#1}%
\providecommand \enquote  [1]{``#1''}%
\providecommand \bibnamefont  [1]{#1}%
\providecommand \bibfnamefont [1]{#1}%
\providecommand \citenamefont [1]{#1}%
\providecommand \href@noop [0]{\@secondoftwo}%
\providecommand \href [0]{\begingroup \@sanitize@url \@href}%
\providecommand \@href[1]{\@@startlink{#1}\@@href}%
\providecommand \@@href[1]{\endgroup#1\@@endlink}%
\providecommand \@sanitize@url [0]{\catcode `\\12\catcode `\$12\catcode
  `\&12\catcode `\#12\catcode `\^12\catcode `\_12\catcode `\%12\relax}%
\providecommand \@@startlink[1]{}%
\providecommand \@@endlink[0]{}%
\providecommand \url  [0]{\begingroup\@sanitize@url \@url }%
\providecommand \@url [1]{\endgroup\@href {#1}{\urlprefix }}%
\providecommand \urlprefix  [0]{URL }%
\providecommand \Eprint [0]{\href }%
\providecommand \doibase [0]{https://doi.org/}%
\providecommand \selectlanguage [0]{\@gobble}%
\providecommand \bibinfo  [0]{\@secondoftwo}%
\providecommand \bibfield  [0]{\@secondoftwo}%
\providecommand \translation [1]{[#1]}%
\providecommand \BibitemOpen [0]{}%
\providecommand \bibitemStop [0]{}%
\providecommand \bibitemNoStop [0]{.\EOS\space}%
\providecommand \EOS [0]{\spacefactor3000\relax}%
\providecommand \BibitemShut  [1]{\csname bibitem#1\endcsname}%
\let\auto@bib@innerbib\@empty
\bibitem [{\citenamefont {H{\"u}bener}\ \emph {et~al.}(2021)\citenamefont
  {H{\"u}bener}, \citenamefont {De~Giovannini}, \citenamefont {Sch{\"a}fer},
  \citenamefont {Andberger}, \citenamefont {Ruggenthaler}, \citenamefont
  {Faist},\ and\ \citenamefont {Rubio}}]{hubener2021engineering}%
  \BibitemOpen
  \bibfield  {author} {\bibinfo {author} {\bibfnamefont {H.}~\bibnamefont
  {H{\"u}bener}}, \bibinfo {author} {\bibfnamefont {U.}~\bibnamefont
  {De~Giovannini}}, \bibinfo {author} {\bibfnamefont {C.}~\bibnamefont
  {Sch{\"a}fer}}, \bibinfo {author} {\bibfnamefont {J.}~\bibnamefont
  {Andberger}}, \bibinfo {author} {\bibfnamefont {M.}~\bibnamefont
  {Ruggenthaler}}, \bibinfo {author} {\bibfnamefont {J.}~\bibnamefont
  {Faist}},\ and\ \bibinfo {author} {\bibfnamefont {A.}~\bibnamefont {Rubio}},\
  }\href@noop {} {\bibfield  {journal} {\bibinfo  {journal} {Nature materials}\
  }\textbf {\bibinfo {volume} {20}},\ \bibinfo {pages} {438} (\bibinfo {year}
  {2021})}\BibitemShut {NoStop}%
\bibitem [{\citenamefont {Schlawin}\ \emph {et~al.}(2022)\citenamefont
  {Schlawin}, \citenamefont {Kennes},\ and\ \citenamefont
  {Sentef}}]{schlawin2022cavity}%
  \BibitemOpen
  \bibfield  {author} {\bibinfo {author} {\bibfnamefont {F.}~\bibnamefont
  {Schlawin}}, \bibinfo {author} {\bibfnamefont {D.~M.}\ \bibnamefont
  {Kennes}},\ and\ \bibinfo {author} {\bibfnamefont {M.~A.}\ \bibnamefont
  {Sentef}},\ }\href@noop {} {\bibfield  {journal} {\bibinfo  {journal}
  {Applied Physics Reviews}\ }\textbf {\bibinfo {volume} {9}} (\bibinfo {year}
  {2022})}\BibitemShut {NoStop}%
\bibitem [{\citenamefont {Bloch}\ \emph {et~al.}(2022)\citenamefont {Bloch},
  \citenamefont {Cavalleri}, \citenamefont {Galitski}, \citenamefont {Hafezi},\
  and\ \citenamefont {Rubio}}]{bloch2022strongly}%
  \BibitemOpen
  \bibfield  {author} {\bibinfo {author} {\bibfnamefont {J.}~\bibnamefont
  {Bloch}}, \bibinfo {author} {\bibfnamefont {A.}~\bibnamefont {Cavalleri}},
  \bibinfo {author} {\bibfnamefont {V.}~\bibnamefont {Galitski}}, \bibinfo
  {author} {\bibfnamefont {M.}~\bibnamefont {Hafezi}},\ and\ \bibinfo {author}
  {\bibfnamefont {A.}~\bibnamefont {Rubio}},\ }\href@noop {} {\bibfield
  {journal} {\bibinfo  {journal} {Nature}\ }\textbf {\bibinfo {volume} {606}},\
  \bibinfo {pages} {41} (\bibinfo {year} {2022})}\BibitemShut {NoStop}%
\bibitem [{\citenamefont {Forn-D\'{\i}az}\ \emph {et~al.}(2019)\citenamefont
  {Forn-D\'{\i}az}, \citenamefont {Lamata}, \citenamefont {Rico}, \citenamefont
  {Kono},\ and\ \citenamefont {Solano}}]{RevModPhys.91.025005}%
  \BibitemOpen
  \bibfield  {author} {\bibinfo {author} {\bibfnamefont {P.}~\bibnamefont
  {Forn-D\'{\i}az}}, \bibinfo {author} {\bibfnamefont {L.}~\bibnamefont
  {Lamata}}, \bibinfo {author} {\bibfnamefont {E.}~\bibnamefont {Rico}},
  \bibinfo {author} {\bibfnamefont {J.}~\bibnamefont {Kono}},\ and\ \bibinfo
  {author} {\bibfnamefont {E.}~\bibnamefont {Solano}},\ }\href
  {https://doi.org/10.1103/RevModPhys.91.025005} {\bibfield  {journal}
  {\bibinfo  {journal} {Rev. Mod. Phys.}\ }\textbf {\bibinfo {volume} {91}},\
  \bibinfo {pages} {025005} (\bibinfo {year} {2019})}\BibitemShut {NoStop}%
\bibitem [{\citenamefont {Lu}\ \emph {et~al.}(2025)\citenamefont {Lu},
  \citenamefont {Shin}, \citenamefont {Svendsen}, \citenamefont {Latini},
  \citenamefont {Hübener}, \citenamefont {Ruggenthaler},\ and\ \citenamefont
  {Rubio}}]{lu2025cavityengineeringsolidstatematerials}%
  \BibitemOpen
  \bibfield  {author} {\bibinfo {author} {\bibfnamefont {I.-T.}\ \bibnamefont
  {Lu}}, \bibinfo {author} {\bibfnamefont {D.}~\bibnamefont {Shin}}, \bibinfo
  {author} {\bibfnamefont {M.~K.}\ \bibnamefont {Svendsen}}, \bibinfo {author}
  {\bibfnamefont {S.}~\bibnamefont {Latini}}, \bibinfo {author} {\bibfnamefont
  {H.}~\bibnamefont {Hübener}}, \bibinfo {author} {\bibfnamefont
  {M.}~\bibnamefont {Ruggenthaler}},\ and\ \bibinfo {author} {\bibfnamefont
  {A.}~\bibnamefont {Rubio}},\ }\href {https://arxiv.org/abs/2502.03172}
  {\bibinfo {title} {Cavity engineering of solid-state materials without
  external driving}} (\bibinfo {year} {2025}),\ \Eprint
  {https://arxiv.org/abs/2502.03172} {arXiv:2502.03172 [cond-mat.mtrl-sci]}
  \BibitemShut {NoStop}%
\bibitem [{\citenamefont {Li}\ \emph {et~al.}(2018)\citenamefont {Li},
  \citenamefont {Bamba}, \citenamefont {Zhang}, \citenamefont {Fallahi},
  \citenamefont {Gardner}, \citenamefont {Gao}, \citenamefont {Lou},
  \citenamefont {Yoshioka}, \citenamefont {Manfra},\ and\ \citenamefont
  {Kono}}]{li2018vacuum}%
  \BibitemOpen
  \bibfield  {author} {\bibinfo {author} {\bibfnamefont {X.}~\bibnamefont
  {Li}}, \bibinfo {author} {\bibfnamefont {M.}~\bibnamefont {Bamba}}, \bibinfo
  {author} {\bibfnamefont {Q.}~\bibnamefont {Zhang}}, \bibinfo {author}
  {\bibfnamefont {S.}~\bibnamefont {Fallahi}}, \bibinfo {author} {\bibfnamefont
  {G.~C.}\ \bibnamefont {Gardner}}, \bibinfo {author} {\bibfnamefont
  {W.}~\bibnamefont {Gao}}, \bibinfo {author} {\bibfnamefont {M.}~\bibnamefont
  {Lou}}, \bibinfo {author} {\bibfnamefont {K.}~\bibnamefont {Yoshioka}},
  \bibinfo {author} {\bibfnamefont {M.~J.}\ \bibnamefont {Manfra}},\ and\
  \bibinfo {author} {\bibfnamefont {J.}~\bibnamefont {Kono}},\ }\href@noop {}
  {\bibfield  {journal} {\bibinfo  {journal} {Nature Photonics}\ }\textbf
  {\bibinfo {volume} {12}},\ \bibinfo {pages} {324} (\bibinfo {year}
  {2018})}\BibitemShut {NoStop}%
\bibitem [{\citenamefont {Paravicini-Bagliani}\ \emph
  {et~al.}(2019)\citenamefont {Paravicini-Bagliani}, \citenamefont
  {Appugliese}, \citenamefont {Richter}, \citenamefont {Valmorra},
  \citenamefont {Keller}, \citenamefont {Beck}, \citenamefont {Bartolo},
  \citenamefont {R{\"o}ssler}, \citenamefont {Ihn}, \citenamefont {Ensslin}
  \emph {et~al.}}]{paravicini2019magneto}%
  \BibitemOpen
  \bibfield  {author} {\bibinfo {author} {\bibfnamefont {G.~L.}\ \bibnamefont
  {Paravicini-Bagliani}}, \bibinfo {author} {\bibfnamefont {F.}~\bibnamefont
  {Appugliese}}, \bibinfo {author} {\bibfnamefont {E.}~\bibnamefont {Richter}},
  \bibinfo {author} {\bibfnamefont {F.}~\bibnamefont {Valmorra}}, \bibinfo
  {author} {\bibfnamefont {J.}~\bibnamefont {Keller}}, \bibinfo {author}
  {\bibfnamefont {M.}~\bibnamefont {Beck}}, \bibinfo {author} {\bibfnamefont
  {N.}~\bibnamefont {Bartolo}}, \bibinfo {author} {\bibfnamefont
  {C.}~\bibnamefont {R{\"o}ssler}}, \bibinfo {author} {\bibfnamefont
  {T.}~\bibnamefont {Ihn}}, \bibinfo {author} {\bibfnamefont {K.}~\bibnamefont
  {Ensslin}}, \emph {et~al.},\ }\href@noop {} {\bibfield  {journal} {\bibinfo
  {journal} {Nature Physics}\ }\textbf {\bibinfo {volume} {15}},\ \bibinfo
  {pages} {186} (\bibinfo {year} {2019})}\BibitemShut {NoStop}%
\bibitem [{\citenamefont {Thomas}\ \emph {et~al.}(2021)\citenamefont {Thomas},
  \citenamefont {Devaux}, \citenamefont {Nagarajan}, \citenamefont {Rogez},
  \citenamefont {Seidel}, \citenamefont {Richard}, \citenamefont {Genet},
  \citenamefont {Drillon},\ and\ \citenamefont {Ebbesen}}]{thomas2021large}%
  \BibitemOpen
  \bibfield  {author} {\bibinfo {author} {\bibfnamefont {A.}~\bibnamefont
  {Thomas}}, \bibinfo {author} {\bibfnamefont {E.}~\bibnamefont {Devaux}},
  \bibinfo {author} {\bibfnamefont {K.}~\bibnamefont {Nagarajan}}, \bibinfo
  {author} {\bibfnamefont {G.}~\bibnamefont {Rogez}}, \bibinfo {author}
  {\bibfnamefont {M.}~\bibnamefont {Seidel}}, \bibinfo {author} {\bibfnamefont
  {F.}~\bibnamefont {Richard}}, \bibinfo {author} {\bibfnamefont
  {C.}~\bibnamefont {Genet}}, \bibinfo {author} {\bibfnamefont
  {M.}~\bibnamefont {Drillon}},\ and\ \bibinfo {author} {\bibfnamefont {T.~W.}\
  \bibnamefont {Ebbesen}},\ }\href@noop {} {\bibfield  {journal} {\bibinfo
  {journal} {Nano letters}\ }\textbf {\bibinfo {volume} {21}},\ \bibinfo
  {pages} {4365} (\bibinfo {year} {2021})}\BibitemShut {NoStop}%
\bibitem [{\citenamefont {Appugliese}\ \emph {et~al.}(2022)\citenamefont
  {Appugliese}, \citenamefont {Enkner}, \citenamefont {Paravicini-Bagliani},
  \citenamefont {Beck}, \citenamefont {Reichl}, \citenamefont {Wegscheider},
  \citenamefont {Scalari}, \citenamefont {Ciuti},\ and\ \citenamefont
  {Faist}}]{doi:10.1126/science.abl5818}%
  \BibitemOpen
  \bibfield  {author} {\bibinfo {author} {\bibfnamefont {F.}~\bibnamefont
  {Appugliese}}, \bibinfo {author} {\bibfnamefont {J.}~\bibnamefont {Enkner}},
  \bibinfo {author} {\bibfnamefont {G.~L.}\ \bibnamefont
  {Paravicini-Bagliani}}, \bibinfo {author} {\bibfnamefont {M.}~\bibnamefont
  {Beck}}, \bibinfo {author} {\bibfnamefont {C.}~\bibnamefont {Reichl}},
  \bibinfo {author} {\bibfnamefont {W.}~\bibnamefont {Wegscheider}}, \bibinfo
  {author} {\bibfnamefont {G.}~\bibnamefont {Scalari}}, \bibinfo {author}
  {\bibfnamefont {C.}~\bibnamefont {Ciuti}},\ and\ \bibinfo {author}
  {\bibfnamefont {J.}~\bibnamefont {Faist}},\ }\href
  {https://doi.org/10.1126/science.abl5818} {\bibfield  {journal} {\bibinfo
  {journal} {Science}\ }\textbf {\bibinfo {volume} {375}},\ \bibinfo {pages}
  {1030} (\bibinfo {year} {2022})},\ \Eprint
  {https://arxiv.org/abs/https://www.science.org/doi/pdf/10.1126/science.abl5818}
  {https://www.science.org/doi/pdf/10.1126/science.abl5818} \BibitemShut
  {NoStop}%
\bibitem [{\citenamefont {Jarc}\ \emph {et~al.}(2023)\citenamefont {Jarc},
  \citenamefont {Mathengattil}, \citenamefont {Montanaro}, \citenamefont
  {Giusti}, \citenamefont {Rigoni}, \citenamefont {Sergo}, \citenamefont
  {Fassioli}, \citenamefont {Winnerl}, \citenamefont {Dal~Zilio}, \citenamefont
  {Mihailovic}, \citenamefont {Prelov{\v{s}}ek}, \citenamefont {Eckstein},\
  and\ \citenamefont {Fausti}}]{jarc2022cavity}%
  \BibitemOpen
  \bibfield  {author} {\bibinfo {author} {\bibfnamefont {G.}~\bibnamefont
  {Jarc}}, \bibinfo {author} {\bibfnamefont {S.~Y.}\ \bibnamefont
  {Mathengattil}}, \bibinfo {author} {\bibfnamefont {A.}~\bibnamefont
  {Montanaro}}, \bibinfo {author} {\bibfnamefont {F.}~\bibnamefont {Giusti}},
  \bibinfo {author} {\bibfnamefont {E.~M.}\ \bibnamefont {Rigoni}}, \bibinfo
  {author} {\bibfnamefont {R.}~\bibnamefont {Sergo}}, \bibinfo {author}
  {\bibfnamefont {F.}~\bibnamefont {Fassioli}}, \bibinfo {author}
  {\bibfnamefont {S.}~\bibnamefont {Winnerl}}, \bibinfo {author} {\bibfnamefont
  {S.}~\bibnamefont {Dal~Zilio}}, \bibinfo {author} {\bibfnamefont
  {D.}~\bibnamefont {Mihailovic}}, \bibinfo {author} {\bibfnamefont
  {P.}~\bibnamefont {Prelov{\v{s}}ek}}, \bibinfo {author} {\bibfnamefont
  {M.}~\bibnamefont {Eckstein}},\ and\ \bibinfo {author} {\bibfnamefont
  {D.}~\bibnamefont {Fausti}},\ }\href
  {https://doi.org/10.1038/s41586-023-06596-2} {\bibfield  {journal} {\bibinfo
  {journal} {Nature}\ }\textbf {\bibinfo {volume} {622}},\ \bibinfo {pages}
  {487} (\bibinfo {year} {2023})}\BibitemShut {NoStop}%
\bibitem [{\citenamefont {Herzig~Sheinfux}\ \emph {et~al.}(2024)\citenamefont
  {Herzig~Sheinfux}, \citenamefont {Orsini}, \citenamefont {Jung},
  \citenamefont {Torre}, \citenamefont {Ceccanti}, \citenamefont {Marconi},
  \citenamefont {Maniyara}, \citenamefont {Barcons~Ruiz}, \citenamefont
  {H{\"o}tger}, \citenamefont {Bertini} \emph {et~al.}}]{herzig2024high}%
  \BibitemOpen
  \bibfield  {author} {\bibinfo {author} {\bibfnamefont {H.}~\bibnamefont
  {Herzig~Sheinfux}}, \bibinfo {author} {\bibfnamefont {L.}~\bibnamefont
  {Orsini}}, \bibinfo {author} {\bibfnamefont {M.}~\bibnamefont {Jung}},
  \bibinfo {author} {\bibfnamefont {I.}~\bibnamefont {Torre}}, \bibinfo
  {author} {\bibfnamefont {M.}~\bibnamefont {Ceccanti}}, \bibinfo {author}
  {\bibfnamefont {S.}~\bibnamefont {Marconi}}, \bibinfo {author} {\bibfnamefont
  {R.}~\bibnamefont {Maniyara}}, \bibinfo {author} {\bibfnamefont
  {D.}~\bibnamefont {Barcons~Ruiz}}, \bibinfo {author} {\bibfnamefont
  {A.}~\bibnamefont {H{\"o}tger}}, \bibinfo {author} {\bibfnamefont
  {R.}~\bibnamefont {Bertini}}, \emph {et~al.},\ }\href@noop {} {\bibfield
  {journal} {\bibinfo  {journal} {Nature Materials}\ }\textbf {\bibinfo
  {volume} {23}},\ \bibinfo {pages} {499} (\bibinfo {year} {2024})}\BibitemShut
  {NoStop}%
\bibitem [{\citenamefont {Kipp}\ \emph {et~al.}(2024)\citenamefont {Kipp},
  \citenamefont {Bretscher}, \citenamefont {Schulte}, \citenamefont {Herrmann},
  \citenamefont {Kusyak}, \citenamefont {Day}, \citenamefont {Kesavan},
  \citenamefont {Matsuyama}, \citenamefont {Li}, \citenamefont {Langner},
  \citenamefont {Hagelstein}, \citenamefont {Sturm}, \citenamefont {Potts},
  \citenamefont {Eckhardt}, \citenamefont {Huang}, \citenamefont {Watanabe},
  \citenamefont {Taniguchi}, \citenamefont {Rubio}, \citenamefont {Kennes},
  \citenamefont {Sentef}, \citenamefont {Baudin}, \citenamefont {Meier},
  \citenamefont {Michael},\ and\ \citenamefont
  {McIver}}]{kipp2024cavityelectrodynamicsvander}%
  \BibitemOpen
  \bibfield  {author} {\bibinfo {author} {\bibfnamefont {G.}~\bibnamefont
  {Kipp}}, \bibinfo {author} {\bibfnamefont {H.~M.}\ \bibnamefont {Bretscher}},
  \bibinfo {author} {\bibfnamefont {B.}~\bibnamefont {Schulte}}, \bibinfo
  {author} {\bibfnamefont {D.}~\bibnamefont {Herrmann}}, \bibinfo {author}
  {\bibfnamefont {K.}~\bibnamefont {Kusyak}}, \bibinfo {author} {\bibfnamefont
  {M.~W.}\ \bibnamefont {Day}}, \bibinfo {author} {\bibfnamefont
  {S.}~\bibnamefont {Kesavan}}, \bibinfo {author} {\bibfnamefont
  {T.}~\bibnamefont {Matsuyama}}, \bibinfo {author} {\bibfnamefont
  {X.}~\bibnamefont {Li}}, \bibinfo {author} {\bibfnamefont {S.~M.}\
  \bibnamefont {Langner}}, \bibinfo {author} {\bibfnamefont {J.}~\bibnamefont
  {Hagelstein}}, \bibinfo {author} {\bibfnamefont {F.}~\bibnamefont {Sturm}},
  \bibinfo {author} {\bibfnamefont {A.~M.}\ \bibnamefont {Potts}}, \bibinfo
  {author} {\bibfnamefont {C.~J.}\ \bibnamefont {Eckhardt}}, \bibinfo {author}
  {\bibfnamefont {Y.}~\bibnamefont {Huang}}, \bibinfo {author} {\bibfnamefont
  {K.}~\bibnamefont {Watanabe}}, \bibinfo {author} {\bibfnamefont
  {T.}~\bibnamefont {Taniguchi}}, \bibinfo {author} {\bibfnamefont
  {A.}~\bibnamefont {Rubio}}, \bibinfo {author} {\bibfnamefont {D.~M.}\
  \bibnamefont {Kennes}}, \bibinfo {author} {\bibfnamefont {M.~A.}\
  \bibnamefont {Sentef}}, \bibinfo {author} {\bibfnamefont {E.}~\bibnamefont
  {Baudin}}, \bibinfo {author} {\bibfnamefont {G.}~\bibnamefont {Meier}},
  \bibinfo {author} {\bibfnamefont {M.~H.}\ \bibnamefont {Michael}},\ and\
  \bibinfo {author} {\bibfnamefont {J.~W.}\ \bibnamefont {McIver}},\ }\href
  {https://arxiv.org/abs/2403.19745} {\bibinfo {title} {Cavity electrodynamics
  of van der waals heterostructures}} (\bibinfo {year} {2024}),\ \Eprint
  {https://arxiv.org/abs/2403.19745} {arXiv:2403.19745 [cond-mat.mes-hall]}
  \BibitemShut {NoStop}%
\bibitem [{\citenamefont {Novoselov}\ \emph {et~al.}(2016)\citenamefont
  {Novoselov}, \citenamefont {Mishchenko}, \citenamefont {Carvalho},\ and\
  \citenamefont {Castro~Neto}}]{Novoselov_2016}%
  \BibitemOpen
  \bibfield  {author} {\bibinfo {author} {\bibfnamefont {K.~S.}\ \bibnamefont
  {Novoselov}}, \bibinfo {author} {\bibfnamefont {A.}~\bibnamefont
  {Mishchenko}}, \bibinfo {author} {\bibfnamefont {A.}~\bibnamefont
  {Carvalho}},\ and\ \bibinfo {author} {\bibfnamefont {A.~H.}\ \bibnamefont
  {Castro~Neto}},\ }\bibfield  {journal} {\bibinfo  {journal} {Science}\
  }\textbf {\bibinfo {volume} {353}},\ \href
  {https://doi.org/10.1126/science.aac9439} {10.1126/science.aac9439} (\bibinfo
  {year} {2016})\BibitemShut {NoStop}%
\bibitem [{\citenamefont {Li}\ \emph {et~al.}(2021)\citenamefont {Li},
  \citenamefont {Qian},\ and\ \citenamefont {Li}}]{review2d}%
  \BibitemOpen
  \bibfield  {author} {\bibinfo {author} {\bibfnamefont {W.}~\bibnamefont
  {Li}}, \bibinfo {author} {\bibfnamefont {X.}~\bibnamefont {Qian}},\ and\
  \bibinfo {author} {\bibfnamefont {J.}~\bibnamefont {Li}},\ }\href
  {https://www.nature.com/articles/s41578-021-00304-0} {\bibfield  {journal}
  {\bibinfo  {journal} {Nature Reviews Materials}\ } (\bibinfo {year}
  {2021})}\BibitemShut {NoStop}%
\bibitem [{\citenamefont {Ren}\ \emph {et~al.}(2016)\citenamefont {Ren},
  \citenamefont {Qiao},\ and\ \citenamefont {Niu}}]{Ren_2016}%
  \BibitemOpen
  \bibfield  {author} {\bibinfo {author} {\bibfnamefont {Y.}~\bibnamefont
  {Ren}}, \bibinfo {author} {\bibfnamefont {Z.}~\bibnamefont {Qiao}},\ and\
  \bibinfo {author} {\bibfnamefont {Q.}~\bibnamefont {Niu}},\ }\href
  {https://doi.org/10.1088/0034-4885/79/6/066501} {\bibfield  {journal}
  {\bibinfo  {journal} {Reports on Progress in Physics}\ }\textbf {\bibinfo
  {volume} {79}},\ \bibinfo {pages} {066501} (\bibinfo {year}
  {2016})}\BibitemShut {NoStop}%
\bibitem [{\citenamefont {Nguyen}\ \emph
  {et~al.}(2023{\natexlab{a}})\citenamefont {Nguyen}, \citenamefont {Arwas},
  \citenamefont {Lin}, \citenamefont {Yao},\ and\ \citenamefont
  {Ciuti}}]{nguyen2023electronphoton}%
  \BibitemOpen
  \bibfield  {author} {\bibinfo {author} {\bibfnamefont {D.-P.}\ \bibnamefont
  {Nguyen}}, \bibinfo {author} {\bibfnamefont {G.}~\bibnamefont {Arwas}},
  \bibinfo {author} {\bibfnamefont {Z.}~\bibnamefont {Lin}}, \bibinfo {author}
  {\bibfnamefont {W.}~\bibnamefont {Yao}},\ and\ \bibinfo {author}
  {\bibfnamefont {C.}~\bibnamefont {Ciuti}},\ }\href
  {https://doi.org/10.1103/PhysRevLett.131.176602} {\bibfield  {journal}
  {\bibinfo  {journal} {Phys. Rev. Lett.}\ }\textbf {\bibinfo {volume} {131}},\
  \bibinfo {pages} {176602} (\bibinfo {year} {2023}{\natexlab{a}})}\BibitemShut
  {NoStop}%
\bibitem [{\citenamefont {Masuki}\ and\ \citenamefont
  {Ashida}(2024)}]{PhysRevB.109.195173}%
  \BibitemOpen
  \bibfield  {author} {\bibinfo {author} {\bibfnamefont {K.}~\bibnamefont
  {Masuki}}\ and\ \bibinfo {author} {\bibfnamefont {Y.}~\bibnamefont
  {Ashida}},\ }\href {https://doi.org/10.1103/PhysRevB.109.195173} {\bibfield
  {journal} {\bibinfo  {journal} {Phys. Rev. B}\ }\textbf {\bibinfo {volume}
  {109}},\ \bibinfo {pages} {195173} (\bibinfo {year} {2024})}\BibitemShut
  {NoStop}%
\bibitem [{\citenamefont {Jiang}\ \emph {et~al.}(2024)\citenamefont {Jiang},
  \citenamefont {Baggioli},\ and\ \citenamefont
  {Jiang}}]{PhysRevLett.132.166901}%
  \BibitemOpen
  \bibfield  {author} {\bibinfo {author} {\bibfnamefont {C.}~\bibnamefont
  {Jiang}}, \bibinfo {author} {\bibfnamefont {M.}~\bibnamefont {Baggioli}},\
  and\ \bibinfo {author} {\bibfnamefont {Q.-D.}\ \bibnamefont {Jiang}},\ }\href
  {https://doi.org/10.1103/PhysRevLett.132.166901} {\bibfield  {journal}
  {\bibinfo  {journal} {Phys. Rev. Lett.}\ }\textbf {\bibinfo {volume} {132}},\
  \bibinfo {pages} {166901} (\bibinfo {year} {2024})}\BibitemShut {NoStop}%
\bibitem [{\citenamefont {Dag}\ and\ \citenamefont
  {Rokaj}(2024)}]{PhysRevB.110.L121101}%
  \BibitemOpen
  \bibfield  {author} {\bibinfo {author} {\bibfnamefont {C.~B.}\ \bibnamefont
  {Dag}}\ and\ \bibinfo {author} {\bibfnamefont {V.}~\bibnamefont {Rokaj}},\
  }\href {https://doi.org/10.1103/PhysRevB.110.L121101} {\bibfield  {journal}
  {\bibinfo  {journal} {Phys. Rev. B}\ }\textbf {\bibinfo {volume} {110}},\
  \bibinfo {pages} {L121101} (\bibinfo {year} {2024})}\BibitemShut {NoStop}%
\bibitem [{\citenamefont {Tay}\ \emph {et~al.}(2024)\citenamefont {Tay},
  \citenamefont {Sanders}, \citenamefont {Baydin}, \citenamefont {Song},
  \citenamefont {Welakuh}, \citenamefont {Alabastri}, \citenamefont {Rokaj},
  \citenamefont {Dag},\ and\ \citenamefont
  {Kono}}]{tay2024terahertzchiralphotoniccrystalcavities}%
  \BibitemOpen
  \bibfield  {author} {\bibinfo {author} {\bibfnamefont {F.}~\bibnamefont
  {Tay}}, \bibinfo {author} {\bibfnamefont {S.}~\bibnamefont {Sanders}},
  \bibinfo {author} {\bibfnamefont {A.}~\bibnamefont {Baydin}}, \bibinfo
  {author} {\bibfnamefont {Z.}~\bibnamefont {Song}}, \bibinfo {author}
  {\bibfnamefont {D.~M.}\ \bibnamefont {Welakuh}}, \bibinfo {author}
  {\bibfnamefont {A.}~\bibnamefont {Alabastri}}, \bibinfo {author}
  {\bibfnamefont {V.}~\bibnamefont {Rokaj}}, \bibinfo {author} {\bibfnamefont
  {C.~B.}\ \bibnamefont {Dag}},\ and\ \bibinfo {author} {\bibfnamefont
  {J.}~\bibnamefont {Kono}},\ }\href {https://arxiv.org/abs/2410.21171}
  {\bibinfo {title} {Terahertz chiral photonic-crystal cavities with broken
  time-reversal symmetry}} (\bibinfo {year} {2024}),\ \Eprint
  {https://arxiv.org/abs/2410.21171} {arXiv:2410.21171 [physics.optics]}
  \BibitemShut {NoStop}%
\bibitem [{\citenamefont {Su{\'a}rez-Forero}\ \emph {et~al.}(2024)\citenamefont
  {Su{\'a}rez-Forero}, \citenamefont {Ni}, \citenamefont {Sarkar},
  \citenamefont {Jalali~Mehrabad}, \citenamefont {Mechtel}, \citenamefont
  {Simonyan}, \citenamefont {Grankin}, \citenamefont {Watanabe}, \citenamefont
  {Taniguchi}, \citenamefont {Park} \emph {et~al.}}]{suarez2024chiral}%
  \BibitemOpen
  \bibfield  {author} {\bibinfo {author} {\bibfnamefont {D.~G.}\ \bibnamefont
  {Su{\'a}rez-Forero}}, \bibinfo {author} {\bibfnamefont {R.}~\bibnamefont
  {Ni}}, \bibinfo {author} {\bibfnamefont {S.}~\bibnamefont {Sarkar}}, \bibinfo
  {author} {\bibfnamefont {M.}~\bibnamefont {Jalali~Mehrabad}}, \bibinfo
  {author} {\bibfnamefont {E.}~\bibnamefont {Mechtel}}, \bibinfo {author}
  {\bibfnamefont {V.}~\bibnamefont {Simonyan}}, \bibinfo {author}
  {\bibfnamefont {A.}~\bibnamefont {Grankin}}, \bibinfo {author} {\bibfnamefont
  {K.}~\bibnamefont {Watanabe}}, \bibinfo {author} {\bibfnamefont
  {T.}~\bibnamefont {Taniguchi}}, \bibinfo {author} {\bibfnamefont
  {S.}~\bibnamefont {Park}}, \emph {et~al.},\ }\href@noop {} {\bibfield
  {journal} {\bibinfo  {journal} {Science Advances}\ }\textbf {\bibinfo
  {volume} {10}},\ \bibinfo {pages} {eadr5904} (\bibinfo {year}
  {2024})}\BibitemShut {NoStop}%
\bibitem [{\citenamefont {Andberger}\ \emph {et~al.}(2024)\citenamefont
  {Andberger}, \citenamefont {Graziotto}, \citenamefont {Sacchi}, \citenamefont
  {Beck}, \citenamefont {Scalari},\ and\ \citenamefont
  {Faist}}]{PhysRevB.109.L161302}%
  \BibitemOpen
  \bibfield  {author} {\bibinfo {author} {\bibfnamefont {J.}~\bibnamefont
  {Andberger}}, \bibinfo {author} {\bibfnamefont {L.}~\bibnamefont
  {Graziotto}}, \bibinfo {author} {\bibfnamefont {L.}~\bibnamefont {Sacchi}},
  \bibinfo {author} {\bibfnamefont {M.}~\bibnamefont {Beck}}, \bibinfo {author}
  {\bibfnamefont {G.}~\bibnamefont {Scalari}},\ and\ \bibinfo {author}
  {\bibfnamefont {J.}~\bibnamefont {Faist}},\ }\href
  {https://doi.org/10.1103/PhysRevB.109.L161302} {\bibfield  {journal}
  {\bibinfo  {journal} {Phys. Rev. B}\ }\textbf {\bibinfo {volume} {109}},\
  \bibinfo {pages} {L161302} (\bibinfo {year} {2024})}\BibitemShut {NoStop}%
\bibitem [{\citenamefont {Suárez-Forero}\ \emph {et~al.}(2024)\citenamefont
  {Suárez-Forero}, \citenamefont {Mehrabad}, \citenamefont {Vega},
  \citenamefont {González-Tudela},\ and\ \citenamefont
  {Hafezi}}]{suarezforero2024chiralquantumopticsrecent}%
  \BibitemOpen
  \bibfield  {author} {\bibinfo {author} {\bibfnamefont {D.~G.}\ \bibnamefont
  {Suárez-Forero}}, \bibinfo {author} {\bibfnamefont {M.~J.}\ \bibnamefont
  {Mehrabad}}, \bibinfo {author} {\bibfnamefont {C.}~\bibnamefont {Vega}},
  \bibinfo {author} {\bibfnamefont {A.}~\bibnamefont {González-Tudela}},\ and\
  \bibinfo {author} {\bibfnamefont {M.}~\bibnamefont {Hafezi}},\ }\href
  {https://arxiv.org/abs/2411.06495} {\bibinfo {title} {Chiral quantum optics:
  recent developments, and future directions}} (\bibinfo {year} {2024}),\
  \Eprint {https://arxiv.org/abs/2411.06495} {arXiv:2411.06495
  [physics.optics]} \BibitemShut {NoStop}%
\bibitem [{\citenamefont {Wei}\ \emph {et~al.}(2024)\citenamefont {Wei},
  \citenamefont {Yang},\ and\ \citenamefont
  {Jiang}}]{wei2024cavityvacuuminducedchiralspinliquids}%
  \BibitemOpen
  \bibfield  {author} {\bibinfo {author} {\bibfnamefont {C.}~\bibnamefont
  {Wei}}, \bibinfo {author} {\bibfnamefont {L.}~\bibnamefont {Yang}},\ and\
  \bibinfo {author} {\bibfnamefont {Q.-D.}\ \bibnamefont {Jiang}},\ }\href
  {https://arxiv.org/abs/2411.08121} {\bibinfo {title} {Cavity-vacuum-induced
  chiral spin liquids in kagome lattices: Tuning and probing topological
  quantum phases via cavity quantum electrodynamics}} (\bibinfo {year}
  {2024}),\ \Eprint {https://arxiv.org/abs/2411.08121} {arXiv:2411.08121
  [cond-mat.str-el]} \BibitemShut {NoStop}%
\bibitem [{\citenamefont {Kibis}\ \emph {et~al.}(2011)\citenamefont {Kibis},
  \citenamefont {Kyriienko},\ and\ \citenamefont
  {Shelykh}}]{PhysRevB.84.195413}%
  \BibitemOpen
  \bibfield  {author} {\bibinfo {author} {\bibfnamefont {O.~V.}\ \bibnamefont
  {Kibis}}, \bibinfo {author} {\bibfnamefont {O.}~\bibnamefont {Kyriienko}},\
  and\ \bibinfo {author} {\bibfnamefont {I.~A.}\ \bibnamefont {Shelykh}},\
  }\href {https://doi.org/10.1103/PhysRevB.84.195413} {\bibfield  {journal}
  {\bibinfo  {journal} {Phys. Rev. B}\ }\textbf {\bibinfo {volume} {84}},\
  \bibinfo {pages} {195413} (\bibinfo {year} {2011})}\BibitemShut {NoStop}%
\bibitem [{\citenamefont {Wang}\ \emph {et~al.}(2019)\citenamefont {Wang},
  \citenamefont {Ronca},\ and\ \citenamefont {Sentef}}]{PhysRevB.99.235156}%
  \BibitemOpen
  \bibfield  {author} {\bibinfo {author} {\bibfnamefont {X.}~\bibnamefont
  {Wang}}, \bibinfo {author} {\bibfnamefont {E.}~\bibnamefont {Ronca}},\ and\
  \bibinfo {author} {\bibfnamefont {M.~A.}\ \bibnamefont {Sentef}},\ }\href
  {https://doi.org/10.1103/PhysRevB.99.235156} {\bibfield  {journal} {\bibinfo
  {journal} {Phys. Rev. B}\ }\textbf {\bibinfo {volume} {99}},\ \bibinfo
  {pages} {235156} (\bibinfo {year} {2019})}\BibitemShut {NoStop}%
\bibitem [{\citenamefont {Masuki}\ and\ \citenamefont
  {Ashida}(2023)}]{PhysRevB.107.195104}%
  \BibitemOpen
  \bibfield  {author} {\bibinfo {author} {\bibfnamefont {K.}~\bibnamefont
  {Masuki}}\ and\ \bibinfo {author} {\bibfnamefont {Y.}~\bibnamefont
  {Ashida}},\ }\href {https://doi.org/10.1103/PhysRevB.107.195104} {\bibfield
  {journal} {\bibinfo  {journal} {Phys. Rev. B}\ }\textbf {\bibinfo {volume}
  {107}},\ \bibinfo {pages} {195104} (\bibinfo {year} {2023})}\BibitemShut
  {NoStop}%
\bibitem [{sup()}]{supp}%
  \BibitemOpen
  \href@noop {} {\bibinfo  {journal} {See supplementary material for details on
  the MFT, edge modes, double layer graphene and the regime $\omega \sim t$}\
  }\BibitemShut {NoStop}%
\bibitem [{\citenamefont {Digiorgio}\ \emph {et~al.}(2023)\citenamefont
  {Digiorgio}, \citenamefont {Senica}, \citenamefont {Micheletti},
  \citenamefont {Beck}, \citenamefont {Faist},\ and\ \citenamefont
  {Scalari}}]{digiorgio2023surface}%
  \BibitemOpen
\bibfield  {journal} {  }\bibfield  {author} {\bibinfo {author} {\bibfnamefont
  {V.}~\bibnamefont {Digiorgio}}, \bibinfo {author} {\bibfnamefont
  {U.}~\bibnamefont {Senica}}, \bibinfo {author} {\bibfnamefont
  {P.}~\bibnamefont {Micheletti}}, \bibinfo {author} {\bibfnamefont
  {M.}~\bibnamefont {Beck}}, \bibinfo {author} {\bibfnamefont {J.}~\bibnamefont
  {Faist}},\ and\ \bibinfo {author} {\bibfnamefont {G.}~\bibnamefont
  {Scalari}},\ }in\ \href@noop {} {\emph {\bibinfo {booktitle} {EPJ Web of
  Conferences}}},\ Vol.\ \bibinfo {volume} {287}\ (\bibinfo {organization} {EDP
  Sciences},\ \bibinfo {year} {2023})\ p.\ \bibinfo {pages} {07028}\BibitemShut
  {NoStop}%
\bibitem [{\citenamefont {Min}\ and\ \citenamefont
  {MacDonald}(2008)}]{min2008electronic}%
  \BibitemOpen
  \bibfield  {author} {\bibinfo {author} {\bibfnamefont {H.}~\bibnamefont
  {Min}}\ and\ \bibinfo {author} {\bibfnamefont {A.~H.}\ \bibnamefont
  {MacDonald}},\ }\href@noop {} {\bibfield  {journal} {\bibinfo  {journal}
  {Progress of Theoretical Physics Supplement}\ }\textbf {\bibinfo {volume}
  {176}},\ \bibinfo {pages} {227} (\bibinfo {year} {2008})}\BibitemShut
  {NoStop}%
\bibitem [{\citenamefont {McCann}\ and\ \citenamefont
  {Koshino}(2013)}]{McCann_2013}%
  \BibitemOpen
  \bibfield  {author} {\bibinfo {author} {\bibfnamefont {E.}~\bibnamefont
  {McCann}}\ and\ \bibinfo {author} {\bibfnamefont {M.}~\bibnamefont
  {Koshino}},\ }\href {https://doi.org/10.1088/0034-4885/76/5/056503}
  {\bibfield  {journal} {\bibinfo  {journal} {Reports on Progress in Physics}\
  }\textbf {\bibinfo {volume} {76}},\ \bibinfo {pages} {056503} (\bibinfo
  {year} {2013})}\BibitemShut {NoStop}%
\bibitem [{\citenamefont {Nguyen}\ \emph
  {et~al.}(2023{\natexlab{b}})\citenamefont {Nguyen}, \citenamefont {Arwas},
  \citenamefont {Lin}, \citenamefont {Yao},\ and\ \citenamefont
  {Ciuti}}]{PhysRevLett.131.176602}%
  \BibitemOpen
  \bibfield  {author} {\bibinfo {author} {\bibfnamefont {D.-P.}\ \bibnamefont
  {Nguyen}}, \bibinfo {author} {\bibfnamefont {G.}~\bibnamefont {Arwas}},
  \bibinfo {author} {\bibfnamefont {Z.}~\bibnamefont {Lin}}, \bibinfo {author}
  {\bibfnamefont {W.}~\bibnamefont {Yao}},\ and\ \bibinfo {author}
  {\bibfnamefont {C.}~\bibnamefont {Ciuti}},\ }\href
  {https://doi.org/10.1103/PhysRevLett.131.176602} {\bibfield  {journal}
  {\bibinfo  {journal} {Phys. Rev. Lett.}\ }\textbf {\bibinfo {volume} {131}},\
  \bibinfo {pages} {176602} (\bibinfo {year} {2023}{\natexlab{b}})}\BibitemShut
  {NoStop}%
\bibitem [{\citenamefont {Ghorashi}\ \emph {et~al.}(2023)\citenamefont
  {Ghorashi}, \citenamefont {Dunbrack}, \citenamefont {Abouelkomsan},
  \citenamefont {Sun}, \citenamefont {Du},\ and\ \citenamefont
  {Cano}}]{PhysRevLett.130.196201}%
  \BibitemOpen
  \bibfield  {author} {\bibinfo {author} {\bibfnamefont {S.~A.~A.}\
  \bibnamefont {Ghorashi}}, \bibinfo {author} {\bibfnamefont {A.}~\bibnamefont
  {Dunbrack}}, \bibinfo {author} {\bibfnamefont {A.}~\bibnamefont
  {Abouelkomsan}}, \bibinfo {author} {\bibfnamefont {J.}~\bibnamefont {Sun}},
  \bibinfo {author} {\bibfnamefont {X.}~\bibnamefont {Du}},\ and\ \bibinfo
  {author} {\bibfnamefont {J.}~\bibnamefont {Cano}},\ }\href
  {https://doi.org/10.1103/PhysRevLett.130.196201} {\bibfield  {journal}
  {\bibinfo  {journal} {Phys. Rev. Lett.}\ }\textbf {\bibinfo {volume} {130}},\
  \bibinfo {pages} {196201} (\bibinfo {year} {2023})}\BibitemShut {NoStop}%
\bibitem [{\citenamefont {Ghorashi}\ and\ \citenamefont
  {Cano}(2023)}]{PhysRevB.107.195423}%
  \BibitemOpen
  \bibfield  {author} {\bibinfo {author} {\bibfnamefont {S.~A.~A.}\
  \bibnamefont {Ghorashi}}\ and\ \bibinfo {author} {\bibfnamefont
  {J.}~\bibnamefont {Cano}},\ }\href
  {https://doi.org/10.1103/PhysRevB.107.195423} {\bibfield  {journal} {\bibinfo
   {journal} {Phys. Rev. B}\ }\textbf {\bibinfo {volume} {107}},\ \bibinfo
  {pages} {195423} (\bibinfo {year} {2023})}\BibitemShut {NoStop}%
\end{thebibliography}%

\newpage \clearpage

\onecolumngrid
\setcounter{secnumdepth}{3}
\appendix

\begin{center}
	{\large
Tunable Topological Phases in Multilayer Graphene Coupled to a Chiral Cavity
	\vspace{4pt}
	\\
	SUPPLEMENTAL MATERIAL
	}
\end{center}

\section{Explicit expressions for the Light-matter interaction in bilayer graphene \label{sec:symmetries}}

In the main text, we consider a cavity with a single polarization.
Here we derive the general case with both polarizations.
We obtain the cavity-matter interaction by minimally coupling the cavity vector potential to bilayer graphene ($\hbar=1$),
\begin{equation}
    \hat H_{BLG}(\mathbf{k})  =\hat  \psi^{\dagger}(\mathbf{k}) v_{\rm F} \tau^0 (\xi (k_x-\hat A_x) \sigma^1 +(k_y-\hat A_y) \sigma^2)+\frac{t}{2}(\tau^1\sigma^1-\tau^2\sigma^2) \hat \psi(\mathbf{k}) + \sum_{\lambda = R,L} \omega_{\lambda}\left(\hat a^{\dagger}_{\lambda}\hat a_{\lambda}+\frac{1}{2}\right)
    \label{HBLG_C}
\end{equation}
where the full vector potential with both circular polarizations reads
\begin{equation}
    {\mathbf{A}}=\sqrt{\frac{1}{\epsilon_0\mathcal{V}_{\lambda}2\omega_{\lambda}}}\left[\mathbf{e}_R \hat {a}_L+\mathbf{e}_R \hat {a}^{\dagger}_R+\mathbf{e}_L \hat {a}_R+\mathbf{e}_L \hat {a}^{\dagger}_L\right]. \label{eq:circVectorPot}
\end{equation}
We define the parameters again, generalized to include the presence of two polarizations: The cavity frequency is given by $\omega_{\lambda }=\sqrt{\omega_c^2+\omega_{D,\lambda}^2}$ and could be different for right and left polarizations due to the diamagnetic frequency $\omega_{D,\lambda}=e/\sqrt{m\epsilon_0 \mathcal{V}_{\lambda}}$ which is introduced during the shift transformation (see the supplementary material of Ref.~\cite{PhysRevB.110.L121101}); $\mathcal{V}_{\lambda}$ are the effective cavity volumes for different polarizations. Eq.~\eqref{HBLG_C} leads the following explicit expression,
\begin{align}\label{int_BLG}
    H_{\rm int}(\mathbf{k}) \underbrace{=}_{\xi=+1}&\, - v_{\rm F} \left( \left[g_R \hat a^{\dagger}_R + g_L \hat a_L\right]\hat c^{\dagger}_{A\mathbf{k}1} \hat c_{B\mathbf{k}1}+\left[g_R \hat a^{\dagger}_R + g_L \hat a_L\right]\hat c^{\dagger}_{A\mathbf{k}2}\hat c_{B\mathbf{k}2}\right) + \rm h.c. \quad\text{or}\cr
    \underbrace{=}_{\xi=-1}&\, v_{\rm F} \left(\left[g_L \hat a^{\dagger}_L + g_R \hat a_R\right]\hat c^{\dagger}_{A\mathbf{k}1}\hat c_{B\mathbf{k}1}+\left[g_L \hat a^{\dagger}_L + g_R \hat a_R\right]\hat c^{\dagger}_{A\mathbf{k}2}\hat c_{B\mathbf{k}2}\right) + \rm h.c.,
\end{align}
where the light-matter interaction strengths are defined by $g_{\lambda} = \frac{\alpha}{m}\sqrt{2\pi/(\mathcal{V}_{\lambda}\hspace{.5mm}\omega_{\lambda}}) > 0$ following the tight-binding model derivation for monolayer graphene in Refs.~\cite{tay2024terahertzchiralphotoniccrystalcavities, PhysRevB.110.L121101}.

When $g_R \neq g_L$, time-reversal symmetry is broken,
as was derived explicitly in Ref.~\cite{PhysRevB.110.L121101} and can be seen from the action of the time-reversal symmetry operator $\mathcal{T}$ on the circularly polarized photon operators,
\begin{equation}
    \mathcal{T}(\hat a_{L,R})=-\hat a_{R,L}\;\; \textrm{and}\;\; \mathcal{T}(\hat a^{\dagger}_{L,R})=-\hat a^{\dagger}_{R,L}.
\end{equation}

We now derive the action of spatial inversion symmetry. The vector potential is odd under inversion symmetry, i.e., $\mathcal{I}(\mathbf{\hat A})=-\mathbf{\hat A}$. Thus, from Eq.~(\ref{eq:circVectorPot}),
\begin{equation}
    \mathcal{I}(\hat a_{\lambda})= -\hat a_{\lambda} \;\; \textrm{and}\;\; \mathcal{I}(\hat a^{\dagger}_{\lambda})= -\hat a^{\dagger}_{\lambda},
\end{equation}
for both linearly and circularly polarized photon operators.
Importantly, having a single polarization, e.g.,~$\mathbf{\hat A}_R$, as defined in the main text, does not preclude invariance under inversion symmetry.

To determine the action of inversion symmetry on the Hamiltonian in Eq.~(\ref{int_BLG}) also requires the action of inversion symmetry on the electron operators, $\mathcal{I}(\hat c_{A\mathbf{k}1}) = \hat c_{B\mathbf{,-k}2}$, which swaps the two valleys (in addition to exchanging sublattice and layer.)
Thus, Eq.~(\ref{int_BLG}) transforms as:
\begin{align}\label{int_BLG}
    \mathcal{I}(H_{\rm int}(\mathbf{k})) =&\, \mathcal{I}(H_{\rm int}(\mathbf{k}),\xi=+1) + \mathcal{I}(H_{\rm int}(\mathbf{k}),\xi=-1)  \notag \\
    =&  v_{\rm F} \left( \left[g_R \hat a^{\dagger}_R + g_L \hat a_L\right]\hat c^{\dagger}_{B\mathbf{,-k}2} \hat c_{A\mathbf{,-k}2}+\left[g_R \hat a^{\dagger}_R + g_L \hat a_L\right]\hat c^{\dagger}_{B\mathbf{,-k}1}\hat c_{A\mathbf{,-k}1}\right) \notag\\
    -& v_{\rm F} \left(\left[g_L \hat a^{\dagger}_L + g_R \hat a_R\right]\hat c^{\dagger}_{B\mathbf{,-k}2}\hat c_{A\mathbf{,-k}2}+\left[g_L \hat a^{\dagger}_L + g_R \hat a_R\right]\hat c^{\dagger}_{B\mathbf{,-k}1}\hat c_{A\mathbf{,-k}1}\right) + \rm h.c. = H_{\rm int}(\mathbf{-k}).
\end{align}
Therefore, the inversion symmetry of the bilayer graphene is preserved under light-matter interaction, even for a single circular polarization. The same is true for monolayer graphene. Analytical \cite{PhysRevB.110.L121101} and numerical MFT confirms this result, yielding $d_3(\mathbf{k})$ and $-d_3(-\mathbf{k})$ at the $\mathbf{K}$ and $\mathbf{K'}$ valleys, respectively, where  $d_3(\mathbf{k})$ is the coefficient of $\sigma_3$, the sublattice degree of freedom, in the effective MFT Hamiltonian.

\section{Details of mean-field theory}

Here we derive the mean-field Hamiltonian (Eq. (4) of the main text) from the cavity-induced interaction (Eq. (3) of the main text), for a single circular polarization. The intralayer terms at the $\mathbf{K}$ valley can be rewritten (dropping constant terms):
\begin{align}
    - \frac{v_F^2 g^2}{\omega}\sum_{\mathbf{k} \mathbf{k'}} c^{\dagger}_{B\mathbf{k'}l}c_{A\mathbf{k'}l}c^{\dagger}_{A\mathbf{k}l}c_{B\mathbf{k}l}=&- \frac{v_F^2 g^2}{\omega}\sum_{\mathbf{k} \mathbf{k'}} \big(\langle c^{\dagger}_{B\mathbf{k'}l}c_{A\mathbf{k'}l}\rangle c^{\dagger}_{A\mathbf{k}l}c_{B\mathbf{k}l}+\langle c^{\dagger}_{A\mathbf{k}l}c_{B\mathbf{k}l}\rangle c^{\dagger}_{B\mathbf{k'}l}c_{A\mathbf{k'}l} \cr
    &+c^{\dagger}_{B\mathbf{k'}l}c_{B\mathbf{k}l} \langle c_{A\mathbf{k'}l}c^{\dagger}_{A\mathbf{k}l} \rangle + \langle c^{\dagger}_{B\mathbf{k'}l}c_{B\mathbf{k}l} \rangle c_{A\mathbf{k'}l}c^{\dagger}_{A\mathbf{k}l} \big)\cr
    =&- \frac{v_F^2 g^2}{\omega}\sum_{\mathbf{k} \mathbf{k'}} \big(\langle c^{\dagger}_{B\mathbf{k'}l}c_{A\mathbf{k'}l}\rangle c^{\dagger}_{A\mathbf{k}l}c_{B\mathbf{k}l}+\langle c^{\dagger}_{A\mathbf{k}l}c_{B\mathbf{k}l}\rangle c^{\dagger}_{B\mathbf{k'}l}c_{A\mathbf{k'}l} \cr
    &+c^{\dagger}_{B\mathbf{k'}l}c_{B\mathbf{k}l} \langle \delta_{k,k'}-c^{\dagger}_{A\mathbf{k}l}c_{A\mathbf{k'}l} \rangle + \langle c^{\dagger}_{B\mathbf{k'}l}c_{B\mathbf{k}l} \rangle (\delta_{k,k'}-c^{\dagger}_{A\mathbf{k}l}c_{A\mathbf{k'}l}) \big)\cr
    =&- \frac{v_F^2 g^2}{\omega}\sum_{\mathbf{k} \mathbf{k'}} \big(\langle c^{\dagger}_{B\mathbf{k'}l}c_{A\mathbf{k'}l}\rangle c^{\dagger}_{A\mathbf{k}l}c_{B\mathbf{k}l}+\langle c^{\dagger}_{A\mathbf{k}l}c_{B\mathbf{k}l}\rangle c^{\dagger}_{B\mathbf{k'}l}c_{A\mathbf{k'}l} \cr
    &+c^{\dagger}_{B\mathbf{k'}l}c_{B\mathbf{k}l} \delta_{k,k'}- c^{\dagger}_{B\mathbf{k'}l}c_{B\mathbf{k}l} \langle c^{\dagger}_{A\mathbf{k}l}c_{A\mathbf{k'}l} \rangle - \langle c^{\dagger}_{B\mathbf{k'}l}c_{B\mathbf{k}l} \rangle c^{\dagger}_{A\mathbf{k}l}c_{A\mathbf{k'}l}) \big).
\end{align}

Similarly, we can write two interlayer interactions as follows,
\begin{align}
    H^{int,1}_{\rm MFT}=- \frac{v_F^2 g^2}{\omega} \sum_{\mathbf{k}\mathbf{k}'} c^{\dagger}_{A\mathbf{k'},2}c_{B\mathbf{k'},2}c^{\dagger}_{B\mathbf{k},1}c_{A\mathbf{k},1} =& - \frac{v_F^2 g^2}{\omega} \sum_{\mathbf{k}\mathbf{k}'} \bigg(\langle c^{\dagger}_{A\mathbf{k}',2}c_{B\mathbf{k}',2}\rangle c^{\dagger}_{B\mathbf{k}1}c_{A\mathbf{k}1}+c^{\dagger}_{A\mathbf{k}',2}c_{B\mathbf{k}',2}\langle c^{\dagger}_{B\mathbf{k}1}c_{A\mathbf{k}1}\rangle \cr
    &-\langle c^{\dagger}_{A\mathbf{k}',2}c_{A\mathbf{k}1}\rangle c^{\dagger}_{B\mathbf{k}1}c_{B\mathbf{k}',2}-c^{\dagger}_{A\mathbf{k}',2}c_{A\mathbf{k}1}\langle c^{\dagger}_{B\mathbf{k}1}c_{B\mathbf{k}',2}\rangle\bigg)\cr
    H^{int,2}_{\rm MFT}=- \frac{v_F^2 g^2}{\omega} \sum_{\mathbf{k}\mathbf{k}'} c^{\dagger}_{A\mathbf{k'},1}c_{B\mathbf{k'},1}c^{\dagger}_{B\mathbf{k},2}c_{A\mathbf{k},2} =& - \frac{v_F^2 g^2}{\omega} \sum_{\mathbf{k}\mathbf{k}'} \bigg(\langle c^{\dagger}_{A\mathbf{k}',1}c_{B\mathbf{k}',1}\rangle c^{\dagger}_{B\mathbf{k}2}c_{A\mathbf{k}2}+c^{\dagger}_{A\mathbf{k}',1}c_{B\mathbf{k}',1}\langle c^{\dagger}_{B\mathbf{k}2}c_{A\mathbf{k}2}\rangle \cr
    &-\langle c^{\dagger}_{A\mathbf{k}',1}c_{A\mathbf{k}2}\rangle c^{\dagger}_{B\mathbf{k}2}c_{B\mathbf{k}',1}-c^{\dagger}_{A\mathbf{k}',1}c_{A\mathbf{k}2}\langle c^{\dagger}_{B\mathbf{k}2}c_{B\mathbf{k}',1}\rangle\bigg),
\end{align}

Therefore, the MFT Hamiltonian at the $\mathbf{K}$ valley is,
\begin{eqnarray}
    H_{\rm MFT}(\mathbf{K})= H_0(\mathbf{K}) + \Delta(\mathbf{K},\mathbf{k}) \label{eq:MFT}
\end{eqnarray}
where the order parameters are
\begin{eqnarray}
  \Delta(\mathbf{K},\mathbf{k}) &=&- \frac{v_F^2 g^2}{\omega}\left[\begin{array}{c|c|c|c}
      -\langle c^{\dagger}_{B\mathbf{k}1}c_{B\mathbf{k}1} \rangle  & \sum_{k'}\langle c^{\dagger}_{B\mathbf{k'}1}c_{A\mathbf{k'}1} \rangle & -\langle c^{\dagger}_{B\mathbf{k}2}c_{B\mathbf{k}1} \rangle & 0 \\ [6pt]
      \hline
      \sum_{k'} \langle c^{\dagger}_{A\mathbf{k'}1}c_{B\mathbf{k'}1} \rangle & -\langle c^{\dagger}_{A\mathbf{k}1}c_{A\mathbf{k}1} \rangle + 1 & 0 & -\langle c^{\dagger}_{A\mathbf{k}2}c_{A\mathbf{k}1} \rangle \\[6pt]
       \hline
       -\langle c^{\dagger}_{B\mathbf{k}1}c_{B\mathbf{k}2} \rangle & 0 & -\langle c^{\dagger}_{B\mathbf{k}2}c_{B\mathbf{k}2} \rangle & \sum_{k'}\langle c^{\dagger}_{B\mathbf{k'}2}c_{A\mathbf{k'}2} \rangle \\[6pt]
       \hline
       0 & -\langle c^{\dagger}_{A\mathbf{k}1}c_{A\mathbf{k}2} \rangle & \sum_{k'} \langle c^{\dagger}_{A\mathbf{k'}2}c_{B\mathbf{k'}2} \rangle & -\langle c^{\dagger}_{A\mathbf{k}2}c_{A\mathbf{k}2} \rangle + 1
  \end{array}\right]
\end{eqnarray}
As explained in the main text, we only study the order parameters on the diagonal because they remain finite when $\mathbf{k}=0$; this yields Eq. (4) of the main text, which
was solved self-consistently to obtain the cavity-induced gap.

To determine whether time reversal and inversion symmetry are broken, we need to compare the two valleys. We write the interaction part of the vacuum projected SW Hamiltonian at the $\mathbf{K'}$ valley as
\begin{eqnarray}
    \mathcal{\hat H}_{\mathbf{K'},\rm vac}(\mathbf{k}) &=&  - \frac{v_F^2 g^2}{\omega}\sum_{\mathbf{k} \mathbf{k'}}\Bigg( \sum_{l} \bigg[ c^{\dagger}_{B\mathbf{k'}l}c_{A\mathbf{k'}l}c^{\dagger}_{A\mathbf{k}l}c_{B\mathbf{k}l} - (n_{B\mathbf{k}l}-n_{A\mathbf{k}l}) \bigg]  + \sum_{m\neq n}\hat c^{\dagger}_{B\mathbf{k'}m}\hat c_{A\mathbf{k'}m}\hat c^{\dagger}_{A\mathbf{k}n}\hat c_{B\mathbf{k}n}\Bigg).
\end{eqnarray}
The only difference compared to the $\mathbf{K}$ valley is the term proportional to $-\tau^0\sigma^3$. The interlayer interactions are exactly the same in both valleys. Hence the MFT order parameters at this valley follow as
\begin{align}
  \Delta(\mathbf{K'},\mathbf{k})=- \frac{v_F^2 g^2}{\omega}\left[\begin{array}{c|c|c|c}
      -\langle c^{\dagger}_{B\mathbf{k}1}c_{B\mathbf{k}1} \rangle +1  & \sum_{k'}\langle c^{\dagger}_{B\mathbf{k'}1}c_{A\mathbf{k'}1} \rangle & -\langle c^{\dagger}_{B\mathbf{k}2}c_{B\mathbf{k}1} \rangle & 0 \\ [6pt]
      \hline
      \sum_{k'} \langle c^{\dagger}_{A\mathbf{k'}1}c_{B\mathbf{k'}1} \rangle & -\langle c^{\dagger}_{A\mathbf{k}1}c_{A\mathbf{k}1} \rangle & 0 & -\langle c^{\dagger}_{A\mathbf{k}2}c_{A\mathbf{k}1} \rangle \\[6pt]
       \hline
       -\langle c^{\dagger}_{B\mathbf{k}1}c_{B\mathbf{k}2} \rangle & 0 & -\langle c^{\dagger}_{B\mathbf{k}2}c_{B\mathbf{k}2} \rangle +1 & \sum_{k'}\langle c^{\dagger}_{B\mathbf{k'}2}c_{A\mathbf{k'}2} \rangle \\[6pt]
       \hline
       0 & -\langle c^{\dagger}_{A\mathbf{k}1}c_{A\mathbf{k}2} \rangle & \sum_{k'} \langle c^{\dagger}_{A\mathbf{k'}2}c_{B\mathbf{k'}2} \rangle & -\langle c^{\dagger}_{A\mathbf{k}2}c_{A\mathbf{k}2} \rangle
  \end{array}\right].
\end{align}

\section{Slab Hamiltonian with zigzag edge} \label{sec:EdgeModes}

\begin{figure}[htb!]
\centering
\includegraphics[width=0.6\columnwidth]{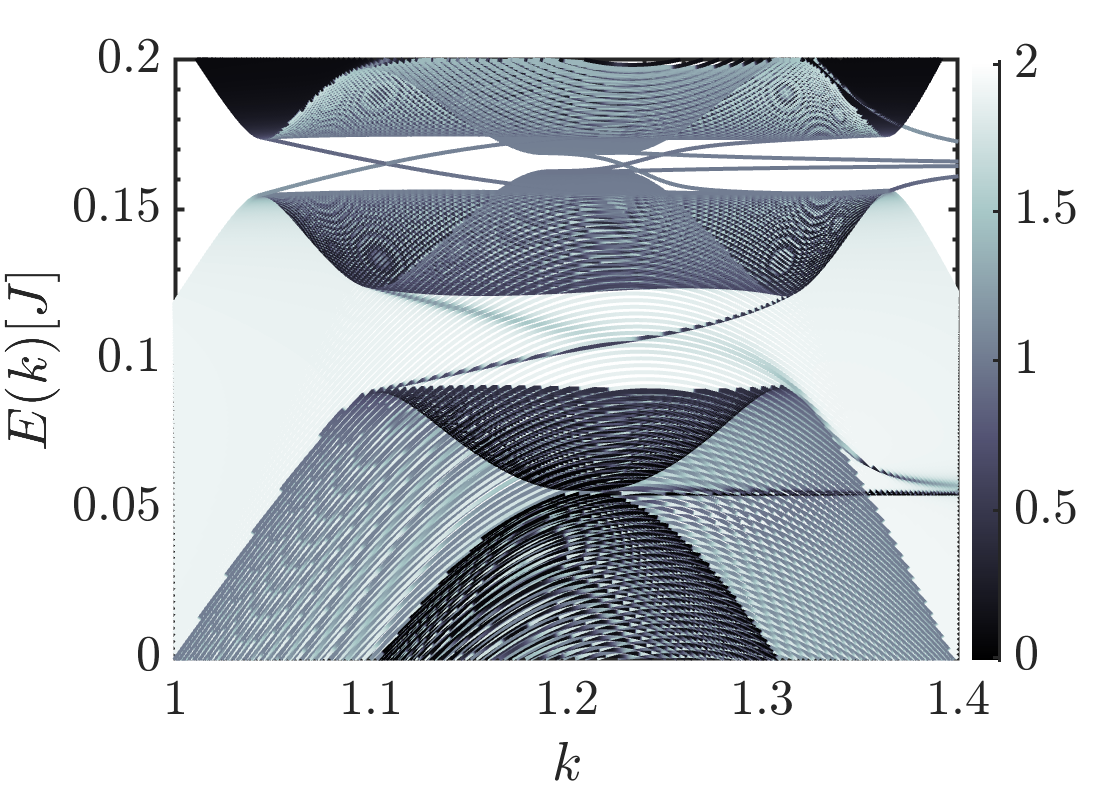}
\caption{Tight-binding band structure for a bilayer graphene slab with zigzag edges, coupled to a chiral cavity mode in the regime $t>\omega$ with $t=0.4J$ and $\omega=0.11J$, where $J=1$ and $\alpha=1$ are fixed for computational convenience.
The x-axis is truncated to a small region around the projection of the $K$ valley.
Colorbar denotes the photon number of the bands. We truncate the photonic Hilbert space to $\langle \hat{n} \rangle = 2$.}\label{SM3}
\end{figure}

We employ the Peierls substitution in the tight-binding model of bilayer graphene \cite{McCann_2013},
\begin{eqnarray}
    H_{\rm tb} (\mathbf k) &=& \sum_{\bf{k},n} \bigg( -J e^{i \mathbf{\delta_n}\cdot \mathbf{A} + i\bf{\delta_n}\cdot \bf{k}} \big[\hat c_{A,1,\bf{k}}^{\dagger} \hat c_{B,1,\bf{k}}+ \hat c_{A,2,\bf{k}}^{\dagger} \hat c_{B,2,\bf{k}} \big] + t \hat c_{A,1,\bf{k}}^{\dagger} \hat c_{B,2,\bf{k}} + \text{h.c.} \bigg),
\end{eqnarray}
where we choose $\boldsymbol{\delta}_1 = \frac{\alpha}{2}\left(\sqrt{3}, 1\right), \,\, \boldsymbol{\delta}_2 = \frac{\alpha}{2}\left(-\sqrt{3}, 1\right), \,\,   \boldsymbol{\delta}_3 =\alpha \left(0,-1\right)$ as the honeycomb bond vectors, and we fix $t=0.4$eV and $J=2.8$eV. Writing the tight-binding Hamiltonian in a mixed momentum-position basis for a zigzag boundary and expanding the Peierls phase to the first order in photon operators results in
\begin{eqnarray}
 \mathcal{H}_{\rm z}&=& -J\sum_{jkl} \biggl[\hat c_{2j,l,k}^{\dagger}\hat c_{2j+1,l,k}
+\hat c_{2j,l,k}^{\dagger}\hat c_{2j-1,l,k} \left(e^{i k a \sqrt{3}/2}+
e^{-i k a \sqrt{3} /2}\right) + \textrm{h.c.} \bigg]+ t \sum_{j,k} \left( \hat c_{2j,k,2}^{\dagger} \hat c_{2j-1,k,1}+\rm h.c. \right)\label{eq:hamiltonianCylinder} \\
&-& iJ g \left(\mathbf{e}_R {a}^{\dagger}+\mathbf{e}_R^* {a} \right) \cdot  \sum_{jkl} \biggl[ \mathbf{\delta_3} \hat c_{2j,l,k}^{\dagger}\hat c_{2j+1,l,k} + \hat c_{2j,l,k}^{\dagger}\hat c_{2j-1,l,k}  \left(\mathbf{\delta_1} e^{i k a \sqrt{3}/2}+
\mathbf{\delta_2} e^{-i k a \sqrt{3} /2}\right) + \textrm{h.c.}  \biggr] + \omega \left(a^{\dagger}a+\frac{1}{2}\right),\notag
\end{eqnarray}
where $j, k, l$ indicate real space position perpendicular to the edge, momentum parallel to the edge, and layer, respectively. 
This Hamiltonian is exactly diagonalized for 1 photon truncation of the photonic Hilbert space in the main text. Here, we show the band structure with 2 photon truncation at a single valley for the same parameter set. We do not observe any changes in the electronic edge modes that connect two valleys at $\braket{\hat n} \approx 0$. However, an important difference compared to the figure in the main text is that the bulk 2-photon bands around $E=0.1J$ hybridize with the 1-photon exchange edge states.
These bulk bands are higher photon excitations, i.e.,~$\braket{\hat n}\approx 2$. In each valley, one of the hybrid edge modes at energy $E=0.1J$ strongly hybridizes with the bulk bands. The fate of these highly hybridized edge modes is left for future research.

\subsection{\label{sec:electronicEdgeModes}Chiral electronic edge modes in the presence of different symmetry breaking perturbations}

The slab band structure in Fig.~4(a) displays symmetry breaking beyond that predicted by the MFT below Eq.~(4).
Here we determine which symmetries are consistent with the bulk and edge band structure in Fig.~4(a) by comparing to the band structure of a slab of bilayer graphene that is not in a cavity.


\begin{figure}[htb!]
\centering
\includegraphics[width=0.7\columnwidth]{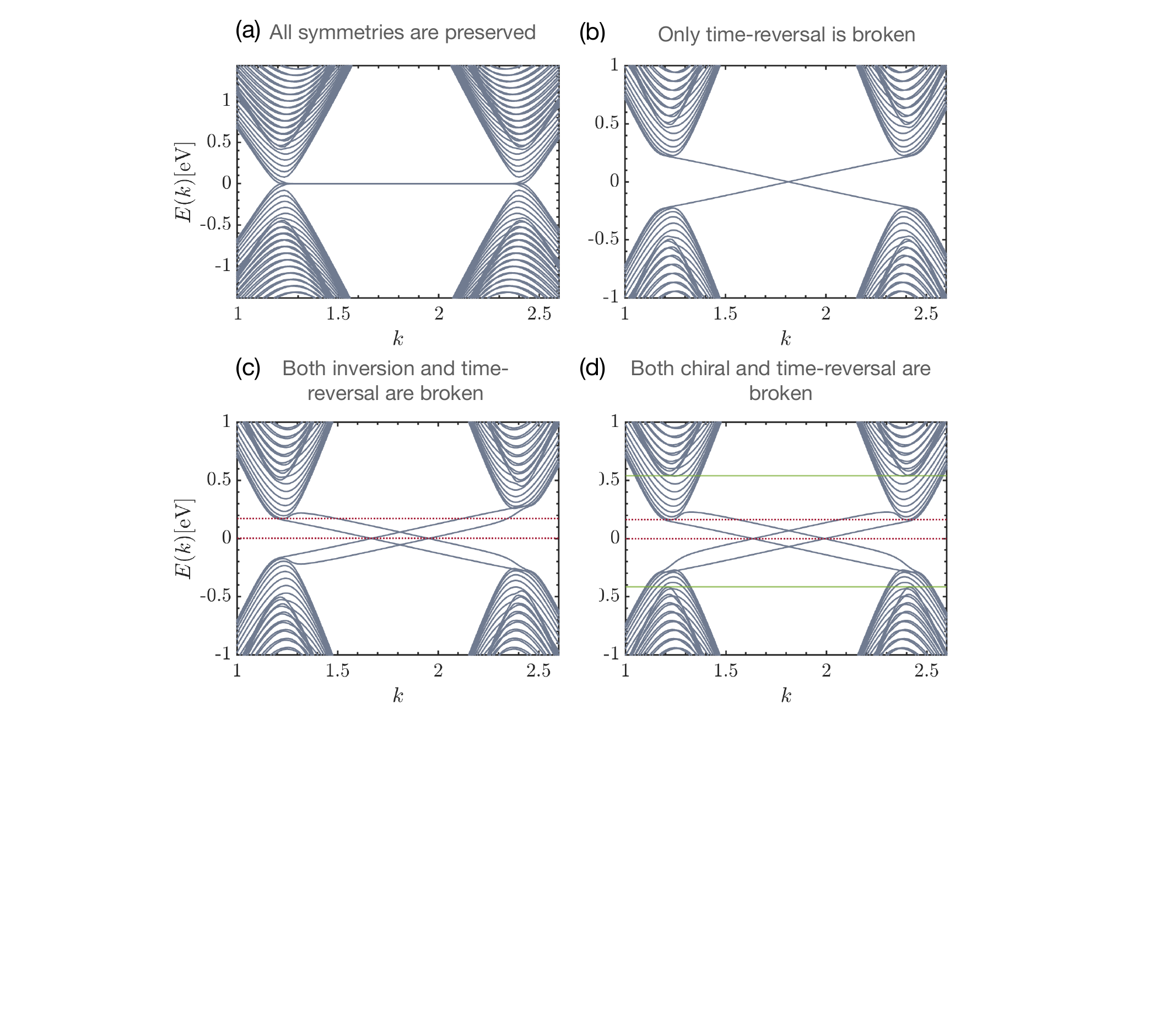}
\caption{The dispersive band structure obtained from Eq.~\eqref{eq:hamiltonianCylinder} in the absence of a cavity, e.g.,~$\omega=g=0$ with $L=100$ unit cells. The tunneling strength $J=2.8$eV and the interlayer tunneling is $t=0.4$eV. Panels (a) through (d) show, respectively, the edge modes when the all symmetries are preserved, only the time-reversal is broken, both inversion and time-reversal are broken, and both chiral symmetry and time-reversal are broken. In (c)-(d) horizontal dotted-red lines are plotted at $E=0$ and at the conduction band minimum in one valley to demonstrate the symmetry breaking between the valleys. In (d), we also plotted solid-green lines to highlight the effect of $\tau_3 \sigma_3$ perturbation on the remote bands. }\label{SM2}
\end{figure}

We diagonalize a slab of bilayer graphene with a zigzag boundary not coupled to a cavity, i.e., the first term in Eq.~\eqref{eq:hamiltonianCylinder}, in the presence of different symmetry breaking terms.
Fig.~\ref{SM2}(a) shows the slab band structure of bilayer graphene when there is no time-reversal or inversion symmetry breaking. There are four degenerate flat bands localized on the zigzag edge.
In (b), we break only the time-reversal symmetry with a term $v_{\rm trs} (\cos(\sqrt{3}k/2+\phi)-\cos(\sqrt{3}k/2-\phi) \tau_0 \sigma_3$ where $v_{\rm trs}=0.1J$ and $\phi=3\sqrt{3/43}$.
The symmetry breaking opens a topological gap in the bulk with Chern number $C=2$. Consequently,
the four flat bands in Fig.~\ref{SM2}(a) split into two dispersing chiral modes on each edge.
The system retains its chiral symmetry, as well as inversion symmetry, which enforces degeneracy between the two valleys.
Further, the combination of chiral and inversion symmetry forces each chiral edge mode to be degenerate.

\begin{figure}[htb!]
\centering
\includegraphics[width=0.9\columnwidth]{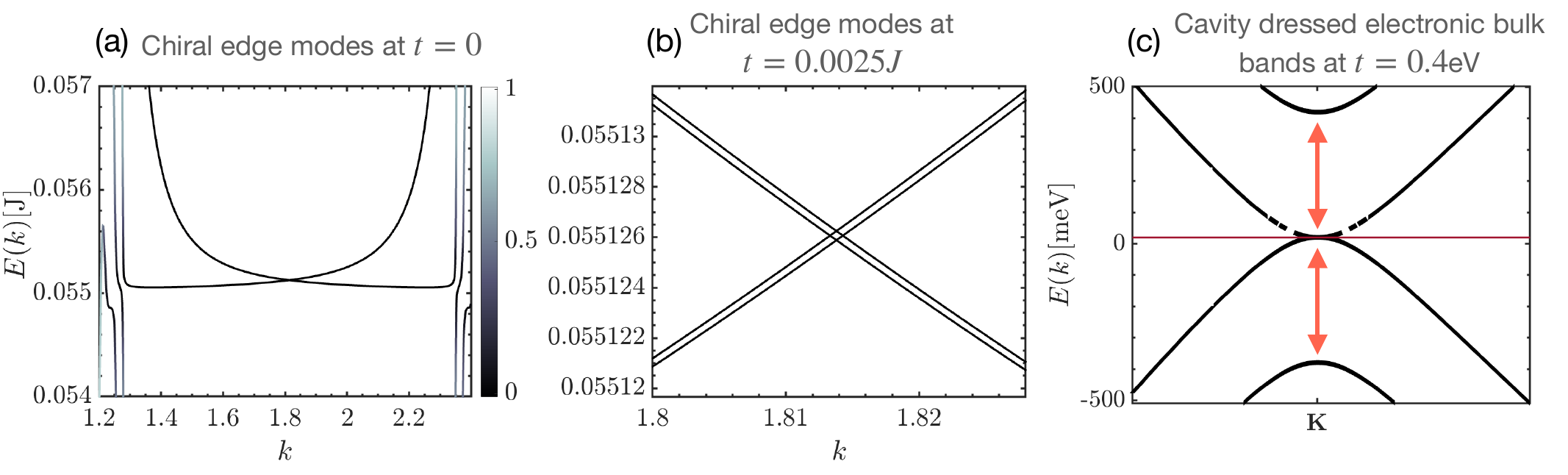}
\caption{(a,b) The dispersive band structure obtained from Eq.~\eqref{eq:hamiltonianCylinder} in the presence of a cavity of frequency $\omega_c=0.11J$ and light-matter coupling strength $g=0.25\omega_c$ with $L=100$ unit cells.
The tunneling strength $J=2.8$eV and the interlayer tunneling is given by (a) $t=0$ and (b) the perturbatively small value of $t=0.0025J$, which is enough to lift the edge state degeneracy. (c) The cavity dressed electronic bulk bands at $t=0.4$eV, $\omega_c=10$THz and $\chi=0.75\times 10^{-5}$ calculated from tight-binding model on torus via the Peierls substitution. The red-solid horizontal line is the cavity zero point energy, which coincides with the energy of the low-energy vacuum band. The energy gaps between low-energy and remote bands shown with red arrows are the same, suggesting that the symmetry breaking perturbation is not $\tau_3 \sigma_3$.}\label{SM5}
\end{figure}

Fig.~\ref{SM2}(c) shows the slab band structure with both inversion and time-reversal symmetry breaking; the inversion breaking term is chosen to be $v_0 \tau_3 \sigma_0$ with $v_0=0.01J$ and $v_{\rm trs}=0.1J$, which could be achieved with a displacement field.
The combination of broken inversion and time-reversal symmetries means that there is no symmetry relating the two valleys, which is evident in the bulk bands in Fig.~\ref{SM2}(c). In addition, the degeneracy of the chiral edge modes is lifted.

In Fig.~\ref{SM2}(d), inversion symmetry is preserved but the chiral symmetry is broken with a term $v_c \tau_3 \sigma_3$ where $v_c=0.025J$ and $v_{\rm trs}=0.1J$.
Here the valleys are degenerate due to the presence of inversion symmetry, but the broken chiral symmetry is evident in both the bulk and surface bands.
Again the chiral edge mode degeneracy is completely split.

We now compare to the band structure of a slab of cavity-coupled bilayer graphene, shown in Fig.~4(a). There the degeneracy of the chiral edge modes is completely split, similar to Fig.~\ref{SM2}(c) or (d).
But since the edge modes in Fig.~4(a) strongly break the chiral symmetry, while the valleys remain symmetric,
we conclude that the cavity-coupled slab breaks chiral symmetry while preserving inversion symmetry.
Indeed, we concluded following Eq.~(\ref{int_BLG}) that inversion symmetry is preserved.

However, the symmetry breaking in the cavity shown in Fig.~4(a) displays differences compared to the term proportional to $\tau_3 \sigma_3$ in Fig.~\ref{SM2}(d).
Specifically, the green lines in Fig.~\ref{SM2}(d) mark the bottom(top) of the first remote conduction(valence) bands, showing that the gaps to these bands are not equal. In contrast, the cavity-dressed electronic bands of bilayer graphene in vacuum in Fig.~\ref{SM5}(c) exhibit equal gaps between low-energy and remote bands. This suggests that $\tau_3 \sigma_3$ is not the right description of the symmetry breaking perturbation.
This is consistent with the MFT, which does not predict the presence of a $\tau_3 \sigma_3$ term.

One might discount the vacuum projected MFT because it does not capture the regime $t > \omega$, as we already discussed in the main text.
However, the splitting between the chiral edge modes is apparent even for perturbatively small $t$ where the MFT is valid (see Figs.~\ref{SM5}(a) and (b).) Thus, we conclude that the absence of a $\tau_3\sigma_3$ term in the MFT does not arise because $t>\omega$, but rather because it is outside the paradigm of $\mathbf{k}$-independent vacuum-projected MFT.

\section{Double layer graphene}

\begin{figure}[htb!]
\centering
\includegraphics[width=0.5\columnwidth]{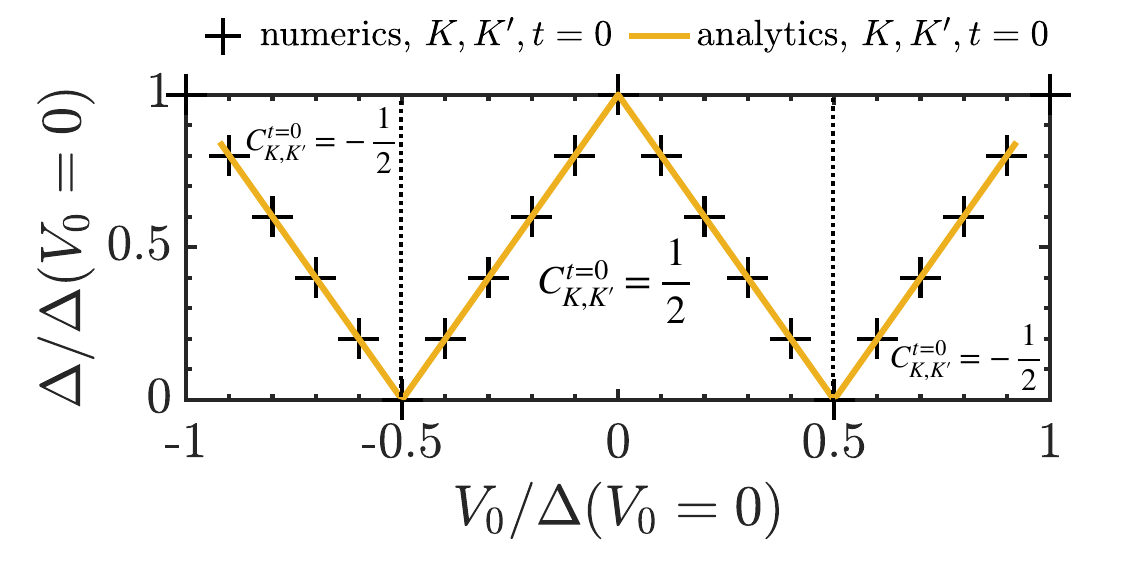}
\caption{The Dirac gap of double-layer graphene with respect to the displacement field where both axes are rescaled by the gap in the absence of the displacement field. The markers are the numerical results obtained from the exact diagonalization of Eq.~(1) in the main text, whereas the yellow solid line is the MFT prediction.}\label{SM1}
\end{figure}
At $t=0$, the gaps at $\mathbf{K}$ and $\mathbf{K'}$ close simultaneously at displacement field $V_0=\pm \Delta(V_0=0)/2$, changing the Chern number of the vacuum band at the Dirac nodes between $C^{t=0}=-1$ and $C^{t=0}=1$. The MFT and the numerics agree, as shown in Fig.~\ref{SM1}.

\section{Electronic band structure in cavity at different interlayer tunneling}

\begin{figure}[htb!]
\centering
\includegraphics[width=0.95\columnwidth]{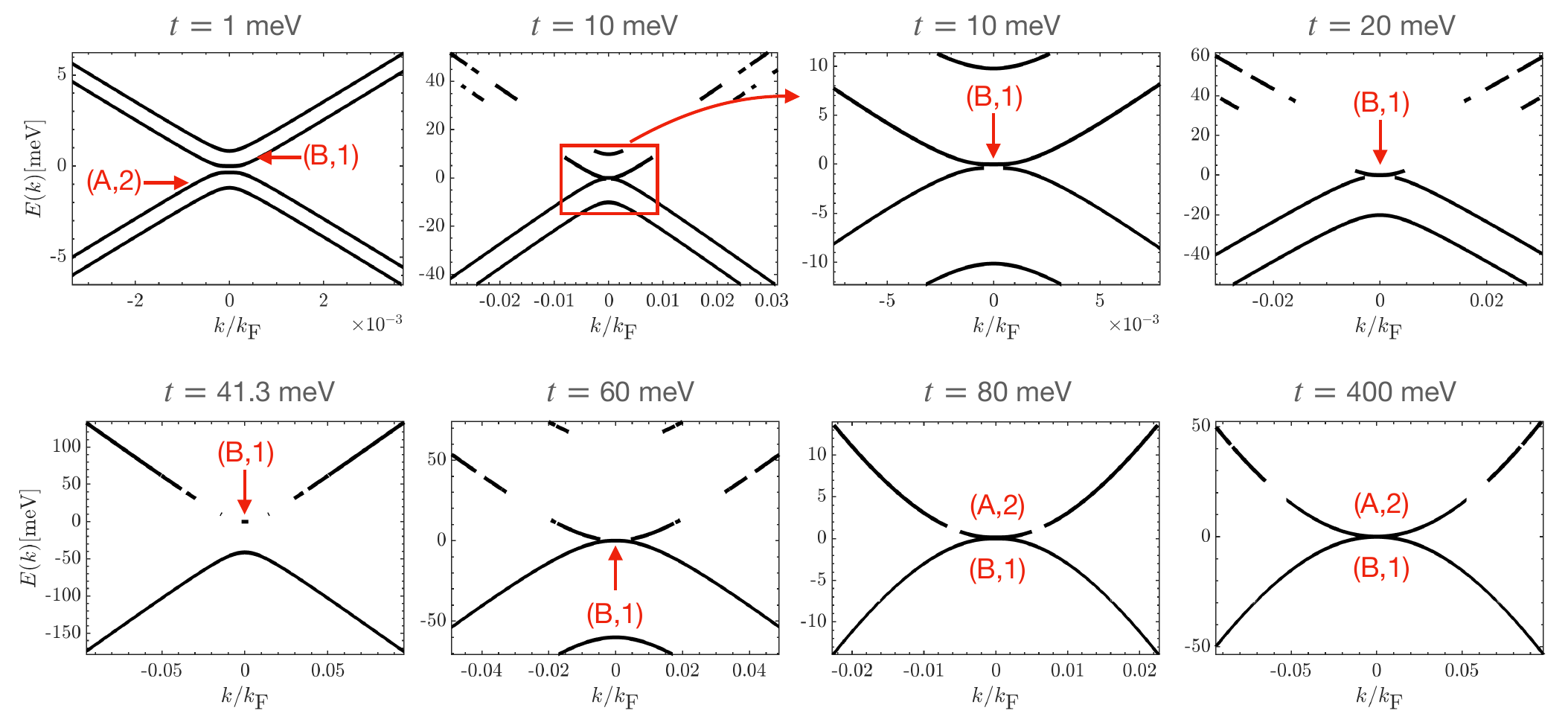}
\caption{The bands with vacuum character $\braket{\hat n} < 0.01$, i.e.,~electronic only, at different interlayer tunneling $t$ in a chiral cavity with frequency $\omega_c=10$ THz and strong light-matter interaction strength $v_{\rm F} g/\omega_c = 0.09$ at zero displacement field. The orbital character of the low-energy bands are marked on the figures.}\label{SM4}
\end{figure}

Here we plot the electronic band structure of bilayer graphene in a cavity in Fig.~\ref{SM4} at different interlayer tunneling values, $t$, and explicitly show why the low-energy band gap is not defined for certain values of $t$.

The lower low-energy band, i.e.,~with $(A,2)$ orbital character at $\mathbf{K}$, strongly hybridizes with the cavity field and ceases to be purely electronic once we increase $t$ from $t=10$ from $t=1$meV. This behavior becomes more pronounced as $t$ increases to $t=20$meV, where we also observe strong hybridization of the upper remote band. At the critical regime of $t\sim \omega$, only two bands survive the cavity hybridization, one is indicated with $(B,1)$ orbital character and the other is the lower remote band with a mixed orbital character between $(A,1)$ and $(B,2)$. Increasing $t$ further changes the curvature of the vacuum band, e.g.,~$t=60$meV, and finally both low-energy bands survive the hybridization at $t=80$meV and beyond. This physics is captured in the main text Fig.~2(c)-(e). Note that one could alternatively define an electronic band gap between the low-energy band with $(B,1)$ orbital character at $\mathbf{K}$ and the lower remote band in this regime $t\sim\omega $, which would be on the order of $\sim 50$meV.

\end{document}